\documentclass{aa}

\usepackage[varg]{txfonts}
\usepackage{siunitx}
\usepackage[version=4]{mhchem}
\usepackage{multirow}
\usepackage{graphicx}
\usepackage[caption=false]{subfig}
\usepackage[strict]{changepage}
\usepackage{xcolor}
\usepackage{ulem}
\usepackage[colorlinks=true,citecolor=blue]{hyperref}
\usepackage{float}
\usepackage{amssymb}
\usepackage{amsmath}
\usepackage{lscape}
\usepackage{mathtools}
\usepackage{natbib}
\usepackage{longtable}
\usepackage{booktabs}
\usepackage{placeins}
                              
\title{Submillimeter observations of molecular gas interacting with the supernova remnant W28}

\author{Parichay Mazumdar\inst{\ref{inst1}}
\and Le Ngoc Tram\inst{\ref{inst1}}
\and Friedrich Wyrowski \inst{\ref{inst1}} 
\and Karl M. Menten\inst{\ref{inst1}} 
\and Xindi Tang\inst{\ref{inst1},\ref{inst2},\ref{inst3}} 
}

\institute{Max-Planck-Institut f\"{u}r Radioastronomie, Auf dem H\"{u}gel 69, 53121 Bonn, Germany \label{inst1} \\
\email{pmazumdar@mpifr-bonn.mpg.de}
\and
Xinjiang Astronomical Observatory, Chinese Academy of Sciences, 830011 Urumqi, PR China \label{inst2}
\and
Key Laboratory of Radio Astronomy, Chinese Academy of Sciences, 830011 Urumqi, PR China \label{inst3}
}

\date{Received xxx / Accepted xxx}

\abstract{Supernovae (SNe) inject large amounts of energy and chemically enriched materials into their surrounding interstellar medium and, in some instances, into molecular clouds {(MCs)}. The interaction of a supernova remnant (SNR) with a MC plays a crucial role in the evolution of the cloud's physical and chemical properties. Despite their importance, only a handful of studies have been made addressing the molecular richness in molecular clouds impacted by  SNRs. (Sub)millimter wavelength observations of MCs affected by SNRs can be used to build a census of their molecular richness, which in turn can motivate various chemical and physical models aimed at explaining the chemical evolution of the clouds.}
{We carried out multi-molecule and multi-transition observations toward the molecular region F {abutting the} SNR W28, containing 1720$\,$MHz OH masers, well-established tracers of SNR-MC interactions. We used the detected lines to constrain the physical conditions of this region.}
{We used the 
APEX Telescope to observe molecular lines in the frequency range $213\rm{-}374\, \textrm{GHz}$. We used non-local thermodynamic equilibrium (non-LTE) RADEX modeling to interpret the observational data.}
{We detected emission from multiple molecular species in the region, namely \ce{CH3OH}, \ce{H2CO}, SO, SiO, CN, CCH, NO, CS, \ce{HCO^+}, HCN, HNC, \ce{N_2H^+}, CO, and from the isotopologues of some of them. We report the first detection of thermally excited (non-maser) \ce{CH3OH} emission toward a SNR. Employing non-LTE RADEX modeling of multiple \ce{H2CO} and \ce{CH3OH} lines, we constrained the kinetic temperature and spatial density in the molecular gas. The gas kinetic temperatures range from 60 to 100$\,$K while the spatial density of the gas ranges from $\num{9e5}$ to $\num{5e6}\,\textrm{cm}^{-3}$. We obtained an ortho-para ratio $\sim$2 for \ce{H2CO}, which indicates that formaldehyde is  most likely formed on dust grain surfaces and not in the gas phase.}
{{Our results show} that molecules as complex as \ce{H2CO} and \ce{CH3OH} can be detected in SNR-MC interactions. {This} could motivate chemical modeling to explore their formation pathways.}

\keywords{ISM: supernova remnants -- ISM: molecules -- astrochemistry -- line: profiles -- molecular processes -- radiative transfer -- shock waves -- ISM: individual objects -- ISM: abundances -- ISM: clouds -- ISM: kinematics and dynamics -- Submillimeter: ISM}

\begin{document}

\titlerunning{A multi-molecular study of W28\,F}
\authorrunning{P. Mazumdar et al.}
\maketitle

\section{Introduction \label{sec:intro}}

    Supernovae (SNe) are highly energetic phenomena injecting large amounts of energy ($\sim$\num{e51} erg) and momentum, along with chemically enriched material into the interstellar medium (ISM) \citep{Dubner2015}. They are considered to be a major source of interstellar turbulence and galactic outflows \citep[and references therein]{li2017}.  
    Their expanding leftovers, referred to as supernova remnants (SNRs), may interact with the ambient or even the parent molecular clouds (MCs) of their massive stellar precursors. This interaction compresses the gas and converts the kinetic energy to thermal energy; thus the gas is heated through the shock passage and the ram pressure of the parent MC is enhanced \citep{chev1999}. All of this triggers chemical reactions which are otherwise unlikely or even impossible in MCs thereby boosting the abundance of molecular species that are usually rare in quiescent gas \citep{Tielens2005}. One example of this is the detection of SiO emission from SNR W51C (\citealt{dumas2014}), a well-established molecular tracer of shocks \citep{schi1997}.
    
    SNR-MC interactions have been intensively studied by observations of the shocked regions by using both molecular and atomic gas tracers (see e.g., \citealt{woo77, den79co, den79oh, woo81, den83, tat85, tat87, fuk, tat90b, tat90a, white87, van, white94, koob,kooa, seta,seta2004, wil, ari, reach, 2000ApJ...544..843R, 2002ApJ...564..302R, 2005ApJ...618..297R, 2007ApJ...664..890N, 2011ApJ...726...76Y, gusdorf, anderl2014, 2019ApJ...884...81R, 2020A&A...644A..64D, 2021arXiv210510617R}).
     Commonly, the rovibrational lines of H$_{2}$ and CO are notably used to characterize the SNR shocks. With no permanent dipole moment, H$_{2}$ is excited at high levels of excitation energy, which is optimally  probed in early regions in shock. Owing to a dipole moment,  
CO complements \ce{H2}  by probing the cooler and denser part in the shocked regions. 
Both the excitation diagram ("Boltzmann plot"), relation between the column density of the rotationally excited levels and their energy above the ground state , e.g., \citealt{2007ApJ...664..890N, 2011ApJ...726...76Y}), and the resolved velocity profiles (e.g., \citealt{2019ApJ...884...81R, 2021arXiv210510617R}) have been  used to constrain the physical properties of the shocked gas.
    
    (Submillimeter and mm observations of SNRs can be used to build a census of their molecular richness, which, in turn, can help in constraining various chemical and physical models aimed at explaining the evolution of MCs interacting with SNRs. These observations also provide input for shocked molecular cloud astrochemistry in general. In one of the few existing studies exploring the molecular richness in an MC-SNR interaction region, \citet{van} carried out a molecular search in the shocked gas region of the SNR IC\,443 and found high frequency lines of \ce{HCO+}, \ce{HCN}, \ce{HNC}, \ce{CS}, \ce{SO}, \ce{SiO}, \ce{H2CO}, and \ce{C2H}. Until today this has been the only extensive study of the inventory of molecular richness existing in SNR-MC interaction regions.
    
    For a long time, maser emission in the 1720\,MHz satellite hyperfine structure transition of the hydroxyl radical (\ce{OH}) has served as a unique molecular signpost of SNR-MC interactions. Even though this line is detected relatively rarely (i.e., in $\sim 10\%$ SNR, \citealt{1997AJ....114.2058G}) it can be used for probing the non-dissociative C-type shock and it provides a direct estimate of the magnetic field strength via the Zeeman effect (\citealt{Lockett1999}).
    
\begin{figure}
            \centering
            \includegraphics[width=0.5\textwidth]{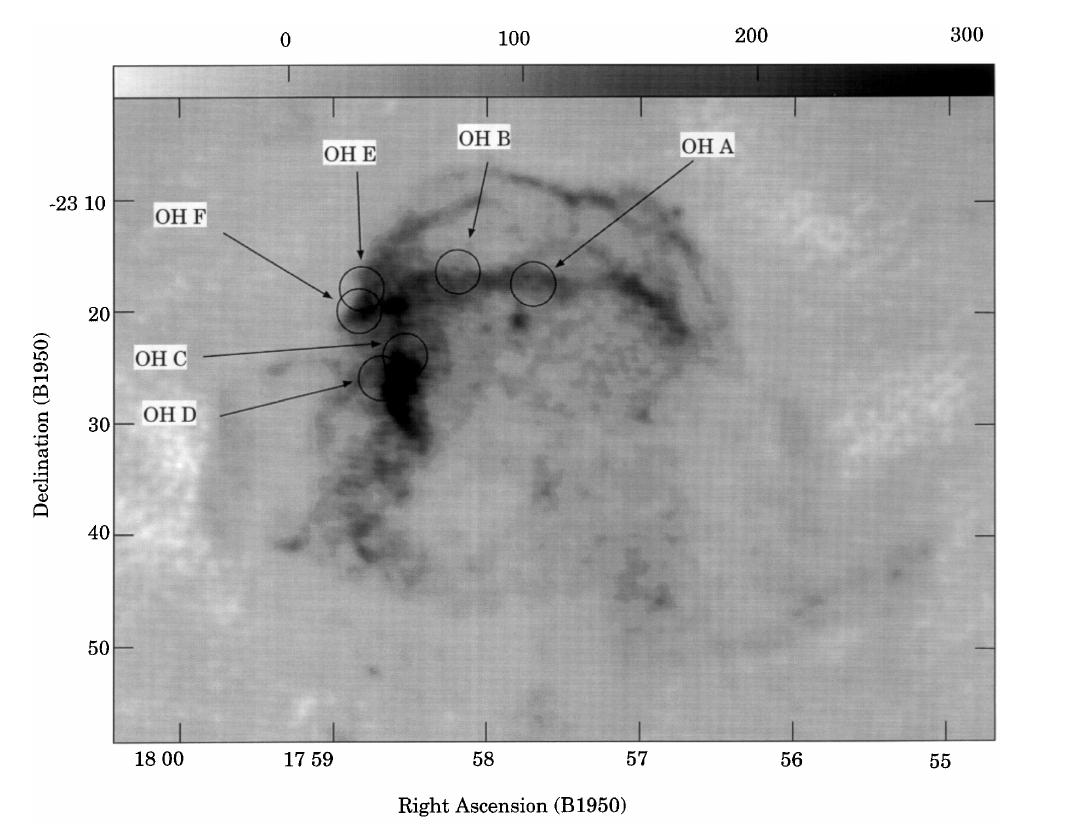}
            \caption{327 MHz radio continuum image \citep{frail93} of the W28 SNR borrowed from Figure 1 of \citet{cla}. The gray scale at the top is in units of mJy beam$^{-1}$. The location of the OH (1720 MHz) emission concentrations have been marked with black circles. The sub-mm wavelength observations reported here {were carried out towards the} F group of OH masers shown in Fig. \ref{fig:map}.
            }
            \label{fig:claussen}
        \end{figure}

    In this work, we report observations with the APEX telescope of emission lines from various molecular species toward region F of the SNR W28, which is a known source of 1720 MHz OH emission 
    (see Figure \ref{fig:claussen}). Our work, thus, fills a gap between (sub)mm studies and FIR/MIR/NIR and radio line studies of  SNR-MC interactions.
    
    W28 is a composite-type SNR that appears as a shell that is $\num{42} \si{\arcminute}$ in diameter in the radio continuum  emission \citep{gre,Dubner2000} and is located at $l$\,=\,6.4$\si{\degree}$, $b$\,=\,--0.2$\si{\degree}$ in the Galactic plane (see Fig. \ref{fig:claussen}). It shows thermal X-Ray emission \citep{Rho2002} as well as gamma ray sources \citep{aharonian2008,giuliani2010,abdo2010}. Many CO studies \citep{ari,reach2005,fukui2008} show massive molecular clouds toward the north-east and the south of the SNR. Emission lines from these clouds were  found to be mostly centered at a local standard of rest velocity, $V_{\text{LSR}}$, 
    of $\approx 7\,\rm km\,s^{-1}$, which is consistent  with values estimated from H{\small{I}} observations \citep{velazquez2002}. The interaction of the SNR shock with the molecular cloud towards its east side appears to be responsible for the enhanced synchrotron and thermal X-ray emissions \citep{Dubner2000}.  Broad \ce{CO} line widths ($\Delta V$\,=\,30--50\,$\si{\kilo\meter\per\second}$) were observed by \cite{frail98} along the gaseous ridges of the molecular cloud facing the shock front while the lines off the ridges were narrow. A very high energy (VHE) $\gamma$-ray emission source, HESS J1801-233 \citep{aharonian2008}, is also coincident with the northeastern boundary of W28, providing further evidence of SNR-MC interactions.
Multiple regions of 1720 MHz OH maser emissions (with $V_{\text{LSR}}$ ranging between 5 and 15$\,\si{\km\per\s}$), loosely distributed in  six areas (A--F) \citep{cla} were also found towards the north eastern regions of W28 (see Fig.\ref{fig:claussen}). These masers were found to be preferentially located along the interface between the shock and the interacting molecular cloud, supporting the hypothesis that they were excited due to the interaction between the two \citep{frail98, Dubner2000}. The APEX observations discussed in the present paper {were carried out towards the} F group of OH masers shown in Fig. \ref{fig:map}.

This paper is structured as follows. We describe the observations and data reduction procedure in Sect. \ref{sec:obs}. Our data analysis and results are described in Sect. \ref{sec:ana}. In Sect. \ref{sec:phys}, {we derive the physical properties} using the non-LTE RADEX code. Our discussion and conclusion are given in Sects. \ref{sec:discuss} and \ref{sec:concl}.
 
\section{Observations and data reduction \label{sec:obs}}

    \subsection{Observations}

        The data were taken with the 12\,m single dish Atacama Pathfinder EXperiment (APEX) submillimeter telescope on Llano Chajnantor, Chile over multiple observing sessions between 2011 August and 2012 July under the project number M-087.F-0041-2011\footnote{\tiny This publication is based on data acquired with the Atacama Pathfinder EXperiment (APEX). APEX is a collaboration between the Max-Planck-Institut f\"{u}r Radioastronomie, the European Southern Observatory, and the Onsala Space Observatory.}. The heterodyne SIS receivers  APEX-1\,(213--275\,$\si{\GHz}$) and FLASH\,(268--374\,$\si{\GHz}$) were used. 
        With FLASH it was possible to observe both side bands simultaneously, while APEX-1 could only observe one side band at a time. System temperatures were in the range 150--400\,K. Table\,\ref{tab:freq} shows a summary of the frequency settings used during the observations.
        Two extended-bandwidth fast Fourier transform spectrometers (XFFTSs) were used as backends. With an instantaneous bandwidth of $\SI{2.5}{\GHz}$, 32768 channels, and a $\SI{1}{\GHz}$ overlap, the backends covered the entire $\SI{4}{\GHz}$ (lower and upper sideband) intermediate-frequency (IF) range of the receivers. This roughly translates to a velocity resolution of 0.094 and 0.072\,$\si{\kilo\meter\per\second}$ at 245 and 320\,GHz, respectively.

\begin{figure}[t]
            \centering
            \includegraphics[width=0.45\textwidth]{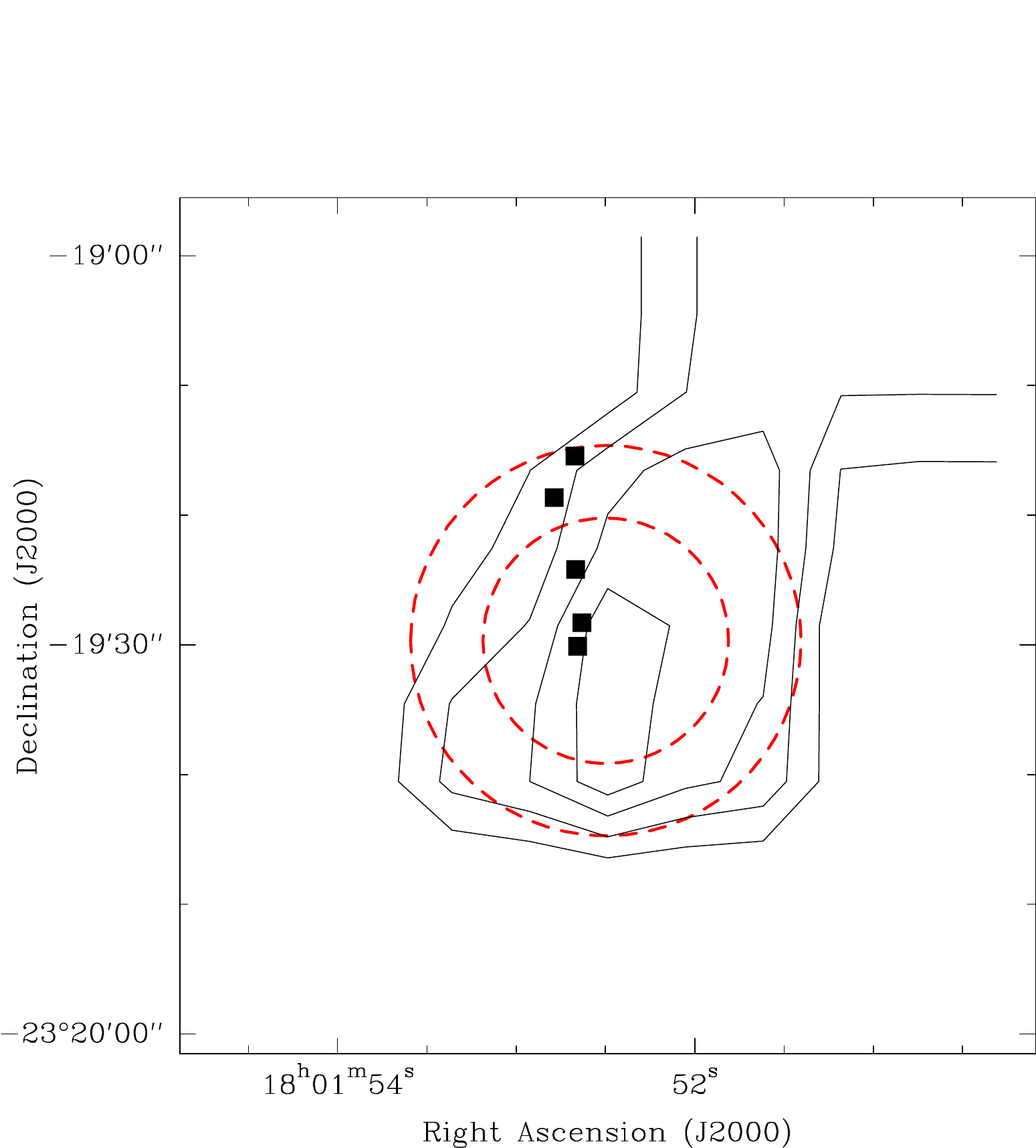}
            \caption{{W28\,F: Map of the main-beam brightness temperature of the \ce{CH3OH}\,($J$\,=\,$6_{0}$--$5_{0}\,A^+$)  transition integrated over the LSR velocity range from --10 to 25\,$\si{\km\per\s}$. The contour values are at three, five, seven and nine times the rms noise level, which is 0.36\,$\si{\kelvin\km\per\s}$}. Overlaid we have the OH maser positions (black filled squares) from \cite{cla} that are located inside the field of view. The large and small circles ($30\si{\arcsecond}$ and $16\si{\arcsecond}$ diameter) represent the FWHM beam width of the APEX telescope at our lowest and highest observing frequency (213 and 374 GHz, respectively). The beams are centered on the methanol peak position (see text) toward  which we performed our long integrations.}
            \label{fig:map}
        \end{figure}

        Guided by the positions of the OH masers in region F, at the beginning of the first observing run, a raster map was observed to find the position of strongest methanol emission in the molecular F region. Figure\,\ref{fig:map} shows an integrated intensity map of the \ce{CH3OH}\,($J$\,=\,$6_{0}$--$5_{0}\,A^+$) transition on a 5\,$\times$\,5 grid (measured with a $10\si{\arcsecond}$ spacing) centered at an offset of $(10\si{\arcsecond},-30\si{\arcsecond})$ from the center of the CO\,$J$\,=\,3--2 map of the W28\,F region presented by \cite{frail98}, which is at 
        $(\alpha,\delta)_{{\rm J}2000}$
        $= 18^{\rm h}01^{\rm m}51\rlap{.}^{\rm s}8$, $\ang{-23;18;59}$. The methanol peak thus is located at $(\alpha,\delta)_{{\rm J}2000}$ $= 18^{\rm h}01^{\rm m}52\rlap{.}^{\rm s}5$, $\ang{-23;19;29}$. All subsequent pointed observations were then carried out towards this peak location of the methanol emission.

        The band pass shape was calibrated using either position switching with $\sim$30 seconds on or off time or in wobbler mode, which observes the reference at a $\SI{90}{\arcsecond}$ offset in azimuth at a rate of $\SI{0.7}{\hertz}$. A frequency dependent beam efficiency was adopted (see Table\,\ref{tab:freq}) for APEX-1 \citep{apex1}. For FLASH, a fixed value of $\eta_{\rm mb}=0.70$ was used. The forward efficiency value of $f_{\rm eff}=0.95$ was used for the whole frequency range. Both of these values for FLASH were based on the information provided on the instrument's webpage. \footnote{\url{https://www.mpifr-bonn.mpg.de/4482467/calibration}}

        \begin{table}[ht!]
            \small
            \centering
            \caption{Summary of frequency setups used during the observations.} \label{tab:freq}
            \begin{tabular} {c c c c c}
                \hline \hline \noalign{\smallskip}
                Instrument &  \parbox [c]{1.2cm}{\centering $\nu$ $(\si{\giga\hertz})$} & \parbox [c] {1.2cm} {\centering $\sigma_{\rm rms}(\si{\kelvin}$)} & {\centering $\eta_{\rm mb}$} & Beam $(\si{\arcsecond})$\\
                \hline \noalign{\smallskip}
                APEX-1          & 213-217      & 0.01 & 0.82 & 30\\
                                & 216-220      & 0.01 & 0.82 & 29.5\\
                                & 220-224      & 0.01 & 0.80 & 29\\
                                & 224-228      & 0.02 & 0.80 & 28.5\\
                                & 234-238      & 0.02 & 0.78 & 27.5\\
                                & 238.5-242.5  & 0.02 & 0.76 & 27\\
                                & 242-246      & 0.01 & 0.76 & 26.5\\
                                & 245-249      & 0.01 & 0.76 & 26\\
                                & 248-252      & 0.02 & 0.74 & 25.5\\
                                & 251-255      & 0.02 & 0.74 & 25\\
                FLASH           & 288.5 -292.5 & 0.01 & 0.70 & 21\\
                                & 292.5-296.5  & 0.02 & 0.70 & 21\\
                                & 296.5-300.5  & 0.01 & 0.70 & 20.5\\
                                & 300.5-304.5  & 0.01 & 0.70 & 20.5\\
                                & 304.5-308.5  & 0.02 & 0.70 & 20\\
                                & 308.5-312.5  & 0.01 & 0.70 & 20\\
                                & 334-338      & 0.02 & 0.70 & 18.5\\
                                & 339-343      & 0.03 & 0.70 & 18\\
                                & 343-347      & 0.02 & 0.70 & 18\\
                                & 346-350      & 0.02 & 0.70 & 17.5\\
                                & 349-353      & 0.02 & 0.70 & 17.5\\
                                & 351-355      & 0.02 & 0.70 & 17.5\\
                                & 355-359      & 0.02 & 0.70 & 17\\
                                & 358-362      & 0.01 & 0.70 & 17\\
                                & 361-365      & 0.02 & 0.70 & 17\\
                                & 370-374      & 0.02 & 0.70 & 16.5\\
                                
                \hline
            \end{tabular}
            \tablefoot{Columns are, left to right, receiver, covered frequency range, rms noise at a velocity resolution of $\sim$1.0\,$\si{\km \per \second}$, main-beam effciency and FWHM beam width.}
        \end{table}

    \subsection{Data reduction}
        
        The data analysis was done using the GILDAS package\footnote{\url{http://www.iram.fr/IRAMFR/GILDAS}}. All the scans with identical backend tuning frequency were stacked channel by channel and then averaged. The frequency ranges of each pair of overlapping backend modules were stitched together and boxcar smoothed to a resolution of $\sim\,1.0\,\si{\km\per\second}$ to provide a single spectrum for the whole 4$\,\si{\giga\hertz}$ receiver bandwidth. Emission lines with peak main-beam brightness temperature, $T_{\rm MB}$, greater than $3\sigma$ at $1.0\,\si{\km\per\second}$ resolution were considered as detection. For each candidate emission line, a baseline was subtracted with 100 channels on either side of the profile. In most of the cases, a first-order baseline was fitted. In about $\sim$20\% of the cases we had to fit a second-order baseline.
        
\section{Analysis and results} \label{sec:ana}
\subsection{Molecules detected in W28$\,$F}

    \begin{figure*}[h!]
      \centering
      \includegraphics[width=1.0\textwidth]{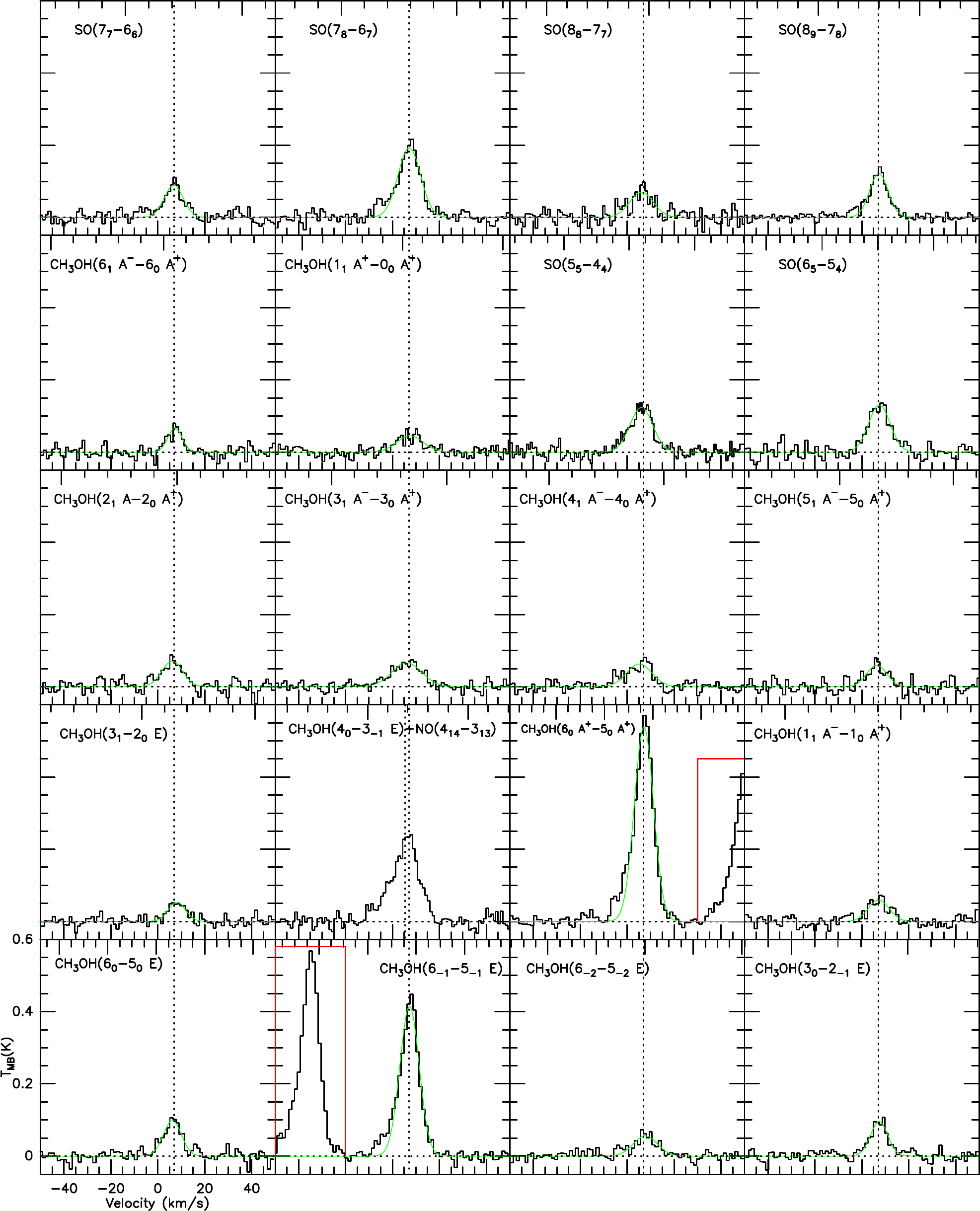}
      \caption{Spectra of the observed species in W28\,F. The intensity is in the $T_{\rm MB}$ scale. The LSR velocity scale in $\si{\km \per \second}$, is appropriate for the line listed in the upper left corner of each panel, except for cases in which more than one line is shown, which are explicitly labelled. Lines marked with red lines are not labelled in the shown window. All the spectra have been smoothed to a velocity resolution of $\approx 1\,\si{\km\per\second}$. Single gaussian fits are shown as green lines overlaid on the spectra. Dashed vertical lines are at an LSR velocity of 7$\,\si{\km \per \second}$ for the labeled lines.}\label{fig:lines-1}
    \end{figure*}
    
    \begin{figure*}[h!]
      \ContinuedFloat
      \centering
      \includegraphics[width=1.0\textwidth]{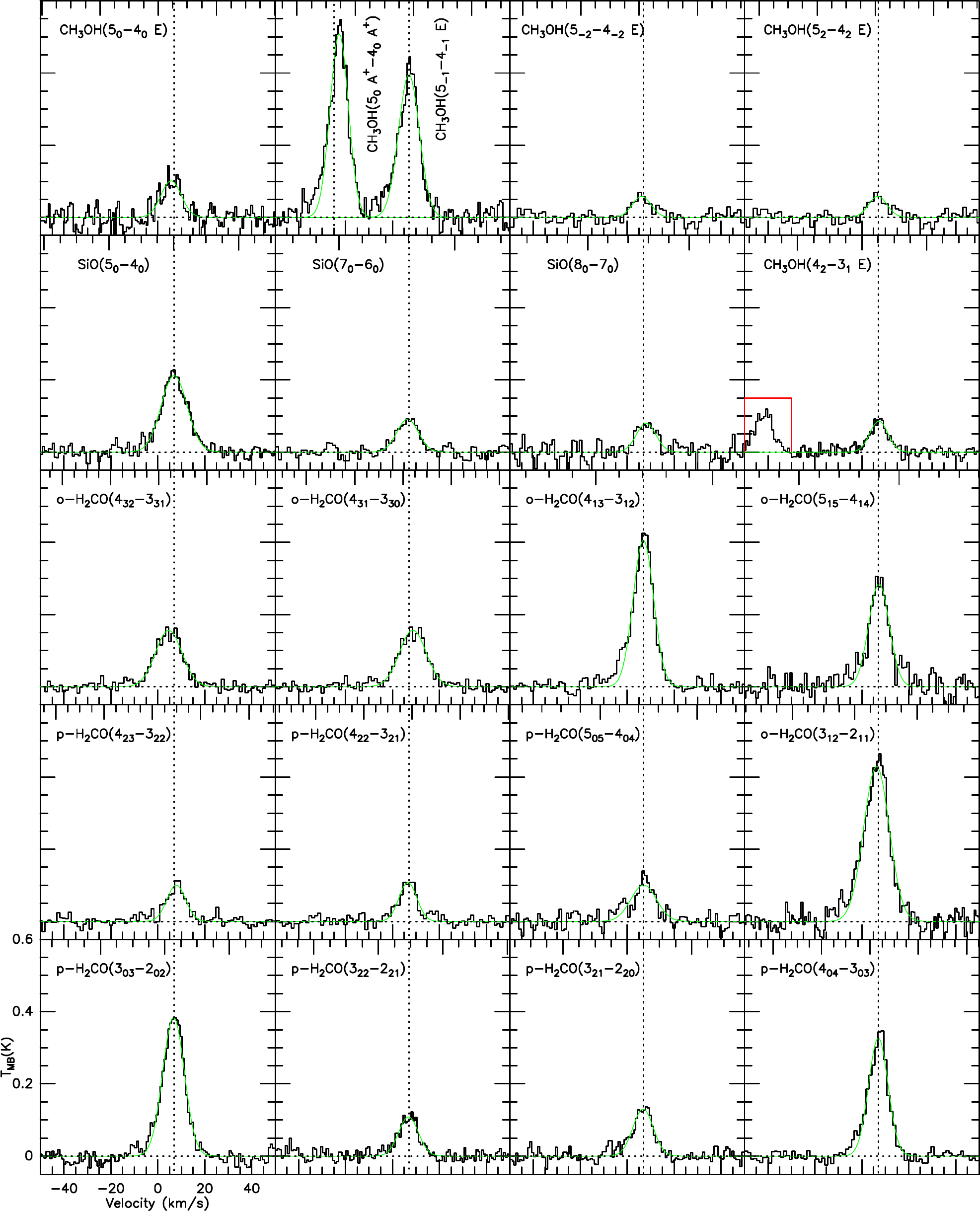}
    \caption{continued .. }
    \end{figure*}
%
%
    \begin{table*}
    \centering
    \caption{Observed molecular species of W28$\,$F and the resulting line parameters from a single Gaussian fit.}

    \begin{tabular}{l l c c c c c c}
    
        \hline \noalign{\smallskip}

        Species & Transition & \parbox[c]{2cm}{\centering Frequency\\$(\si{\MHz})$} & \parbox[c]{2cm}{\centering $E_{\rm up}$\\$(\si{\kelvin})$} & \parbox[c]{2cm} {\centering $\int$$T_{\rm MB}$ d$v$\\ $(\si{\K\km\per\s})$} & \parbox[c]{2cm}{\centering $V_{\rm LSR}$\\$(\si{\km\per\s}$)} & \parbox[c]{2cm}{\centering $\Delta V$\\$(\si{\km\per\s}$)} & \parbox[c]{2cm}{\centering $T_{\rm MB}$\\$(\si{\kelvin}$)} \\

        \hline \noalign{\smallskip}
        \ce{p-H2CO} & $3_{0,3} - 2_{0,2}$ & 218222.192 & 21.0 & 3.94(7) & 6.7(1) & 11.0(2) & 0.34(1) \\ 
        & $3_{2,2} - 2_{2,1}$ & 218475.632 & 68.1 & 0.99(6) & 6.9(3) & 9.9(7) & 0.09(1) \\ 
        & $3_{2,1} - 2_{2,0}$ & 218760.066 & 68.1 & 1.29(7) & 6.8(3) & 10.6(8) & 0.11(1) \\ 
        & $4_{0,4} - 3_{0,3}$ & 290623.405 & 34.9 & 3.34(7) & 7.1(1) & 9.6(2) & 0.33(2) \\ 
        & $4_{2,3} - 3_{2,2}$ & 291237.766 & 82.1 & 1.01(7) & 7.8(3) & 9.4(8) & 0.10(2) \\ 
        & $4_{2,2} - 3_{2,1}$ & 291948.067 & 82.1 & 0.97(7) & 6.7(3) & 8.3(8) & 0.11(2) \\ 
        & $5_{0,5} - 4_{0,4}$ & 362736.048 & 52.3 & 1.3(1) & 7.2(5) & 12(2) & 0.10(2) \\ 
        \ce{o-H2CO} & $3_{1,2} - 2_{1,1}$ & 225697.775 & 33.4 & 4.6(2) & 6.9(2) & 10.7(5) & 0.40(2) \\ 
        & $4_{3,2} - 3_{3,1}$ & 291380.442 & 140.9 & 1.02(6) & - & - & 0.08(2) \tablefootmark{a} \\ 
        & $4_{3,1} - 3_{3,0}$ & 291384.361 & 140.9 & 1.02(6) & - & - & 0.08(2) \tablefootmark{a} \\ 
        & $4_{1,3} - 3_{1,2}$ & 300836.635 & 47.9 & 4.36(8) & 6.96(9) & 10.1(2) & 0.40(2) \\ 
        & $5_{1,5} - 4_{1,4}$ & 351768.645 & 62.5 & 3.2(1) & 7.0(2) & 12.2(5) & 0.25(2) \\ 
        \ce{E-CH3OH} & $4_{2} - 3_{1}\,E$ & 218440.063 & 45.5 & 0.72(1) & 7.2(7) & 9(2) & 0.07(1) \\ 
        & $5_{0} - 4_{0}\,E$ & 241700.159 & 47.9 & 1.0(4) & 6(2) & 10(4) & 0.10(2) \\ 
        & $5_{-1} - 4_{-1}\,E$ & 241767.234 & 40.4 & 4.1(1) & 7.0(1) & 10.5(4) & 0.37(2) \\ 
        & $5_{1} - 4_{1}\,E$ & 241879.025 & 55.9 & - & - & - & $ < 0.06$ \tablefootmark{b} \\ 
        & $5_{-2} - 4_{-2}\,E$ & 241904.147 & 60.7 & 0.34(8) & - & - & 0.03(2) \tablefootmark{a} \\ 
        & $5_{2} - 4_{2}\,E$ & 241904.643 & 57.1 & 0.34(8) & - & - & 0.03(2) \tablefootmark{a} \\ 
        & $6_{0} - 5_{0}\,E$ & 289939.377 & 61.8 & 1.0(1) & 6.3(4) & 10(1) & 0.10(2) \\ 
        & $6_{-1} - 5_{-1}\,E$ & 290069.747 & 54.3 & 4.13(8) & 7.28(9) & 9.5(2) & 0.41(2) \\ 
        & $6_{1} - 5_{1}\,E$ & 290248.685 & 69.8 & - & - & - & $<0.06$ \tablefootmark{b} \\ 
        & $6_{-2} - 5_{-2}\,E$ & 290307.281 & 74.7 & 0.33(6) & - & - & 0.03(1) \tablefootmark{a} \\ 
        & $6_{2} - 5_{2}\,E$ & 290307.738 & 71.0 & 0.33(6) & - & - & 0.33(1) \tablefootmark{a} \\
        & $3_{0} - 2_{-1}\,E$ & 302369.773 & 27.1 & 0.86(7) & 7.4(3) & 8.4(8) & 0.10(2) \\ 
        & $3_{1} - 2_{0}\,E$ & 310192.994 & 35.0 & 0.51(7) & 8.1(6) & 9(1) & 0.05(1) \\ 
        & $4_{0} - 3_{-1}\,E$ & 350687.662 & 36.3 & - & - & - & $\sim 0.07(1)$ \tablefootmark{c}\\ 
        \ce{A-CH3OH} & $5_{0}\,A^+ - 4_{0}\,A^+$ & 241791.352 & 34.8 & 4.8(1) & 7.1(1) & 9.7(2) & 0.47(2) \\ 
        & $5_{2}\,A^- - 4_{2}\,A^-$ & 241842.284 & 72.5 & - & - & - & $ < 0.06$ \tablefootmark{b} \\ 
        & $5_{2}\,A^+ - 4_{2}\,A^+$ & 241887.674 & 72.5 & - & - & - & $ < 0.06$ \tablefootmark{b} \\ 
        & $6_{0}\,A^+ - 5_{0}\,A^+$ & 290110.637 & 48.7 & 5.60(8) & 7.19(6) & 9.6(2) & 0.54(2) \\ 
        & $1_{1}\,A^- - 1_{0}\,A^+$ & 303366.921 & 16.9 & 0.68(9) & 7.2(6) & 10(2) & 0.06(2) \\ 
        & $2_{1}\,A^- - 2_{0}\,A^+$ & 304208.348 & 21.6 & 0.64(8) & 6.2(6) & 9(1) & 0.06(2) \\ 
        & $3_{1}\,A^- - 3_{0}\,A^+$ & 305473.491 & 28.6 & 0.95(7) & 6(1) & 14(2) & 0.06(1) \\ 
        & $4_{1}\,A^- - 4_{0}\,A^+$ & 307165.924 & 38.0 & 0.68(8) & 7.6(4) & 11(1) & 0.08(1) \\ 
        & $5_{1}\,A^- - 5_{0}\,A^+$ & 309290.360 & 49.7 & 0.67(7) & 6(1) & 10(1) & 0.06(2) \\ 
        & $6_{1}\,A^- - 6_{0}\,A^+$ & 311852.612 & 63.7 & 0.58(7) & 6.9(4) & 6.7(9) & 0.08(1) \\ 
        & $1_{1}\,A^+ - 0_{0}\,A^+$ & 350905.100 & 16.8 & 0.70(8) & 7.4(8) & 14(2) & 0.05(1) \\ 
        \ce{SO} & $5_5 - 4_4$ & 215220.653 & 44.1 & 1.30(9) & 6.3(3) & 11.8(7) & 0.11(2) \\ 
        & $6_5 - 5_4$ & 251825.770 & 50.7 & 1.4(1) & 7.4(5) & 10(1) & 0.12(3) \\ 
        & $7_7 - 6_6$ & 301286.124 & 71.0 & 0.80(8) & 6.9(4) & 9(1) & 0.09(2) \\ 
        & $7_8 - 6_7$ & 304077.844 & 62.1 & 2.4(1) & 7.2(2) & 11.8(7) & 0.19(2) \\ 
        & $8_8 - 7_7$ & 344310.612 & 87.5 & 0.8(1) & 6(1) & 10(2) & 0.07(2) \\ 
        & $8_9 - 7_8$ & 346528.481 & 78.8 & 1.4(1) & 6.7(5) & 9(1) & 0.14(1) \\ 
        \ce{SiO} & $5 - 4$ & 217104.919 & 31.3 & 2.45(7) & 7.0(2) & 12.7(4) & 0.18(1) \\ 
        & $7 - 6$ & 303926.960 & 58.3 & 1.01(8) & 6.5(4) & 11(1) & 0.09(2) \\ 
        & $8 - 7$ & 347330.581 & 75.0 & 1.0(1) & 8.1(8) & 14(2) & 0.06(1) \\ 

      \hline \\
    \end{tabular}
    \label{tab:line}
    \tablefoot {{Columns left to right: molecular species, quantum numbers, frequency and upper level energy of transition, velocity-integrated main-beam brightness temperature, centroid LSR velocity, line width, and peak main-beam brightness temperature.} The $T_{\rm MB}$ values are the fit results and not the absolute peak observed values. In cases where no Gaussian fitting could be done, the {observed temperature peaks} have been shown. {The numbers within the parentheses show the error associated with the last significant digit.} \\
    \tablefoottext{a}{These lines were blended with same $E_{\rm up}$ and $A_{\rm ij}$ values. They were fitted with one Gaussian and the integrated intensities were obtained by dividing the result of the fit by the number of blended lines assuming equal contribution from each. No attempt was made to derive LSR velocities or line widths.} \\
    \tablefoottext{b}{$3\sigma$ detection threshold applied for non-detected lines.}\\
    \tablefoottext{c}{{The covered \ce{CH3OH} lines were blended with a \ce{NO} line and hence only an estimate of the upper limit on their peak main-beam brightness temperature could be established. For this, we assumed that the NO lines at $\sim$351050 and $\sim$350690\,$\si{\mega\hertz}$ have similar emission strengths because of their identical $E_{\rm up}$ and very similar $A_{\rm ij}$ values.}} \\
    }
    
    \end{table*}

    The averaged, smoothed, and baseline subtracted spectra were used to identify the spectral lines. For this purpose, we used the JPL spectral line catalog\footnote{\url{https://spec.jpl.nasa.gov/home.html}} \citep{jpl} and the Cologne Database for Molecular Spectroscopy (CDMS) \citep{cdms}. We detected multiple lines of \ce{H2CO}, \ce{CH3OH}, \ce{SO}, \ce{SiO}, \ce{CS}, \ce{CN}, \ce{CCH}, and \ce{NO} as well as many other molecular species including \ce{CO} and its isotopologues, \ce{HCO+}, \ce{H^{13}CO+}, \ce{HCN}, \ce{HNC}, and  \ce{N2H+}. {This is the first detection of thermally excited \ce{CH3OH} emission in an SNR, while methanol maser detections in SNRs have been reported before \citep{philstrom14}.} For some molecules, such as \ce{H2CO} and \ce{CH3OH} there are several transitions close together in frequency space that have been observed simultaneously. This facilitates accurate determination of temperature and densities without significant calibration and beam-filling factor uncertainties. Figure \ref{fig:lines-1} shows some of the detected lines of \ce{CH3OH}, \ce{H2CO}, \ce{SO} and \ce{SiO}. The other detected lines are presented in Fig.\ref{spec}. After identification, a Gaussian profile was fitted to each of them to obtain the following line parameters: local standard of rest (LSR) velocity ($V_{\rm LSR}$), line width ($\Delta V$), peak main-beam temperature ($T_{\rm MB}$), and the integrated intensity ($\int T_{\rm MB}$d$v$). A single Gaussian turned out to be adequate to fit all the spectral lines except for the lines of \ce{CO} and {its isotopologues} which displayed {more complex profiles} including self-absorption. For molecular species with multiple detected lines very closely spaced in frequency, the resulting line profile was non-Gaussian. Thus, no Gaussian fitting was done for such species and only upper limits on their emission were determined from peak $T_{\rm MB}$ values. Table\,\ref{tab:line} lists the spectral lines shown in Fig. \ref{fig:lines-1} with their respective line parameters obtained from the fit. Other molecular line fit parameters are shown in Table. \ref{tab:lines_cont}.

    \subsection{Kinematics}

        \begin{figure}[h!]
            \includegraphics[width=0.48\textwidth]{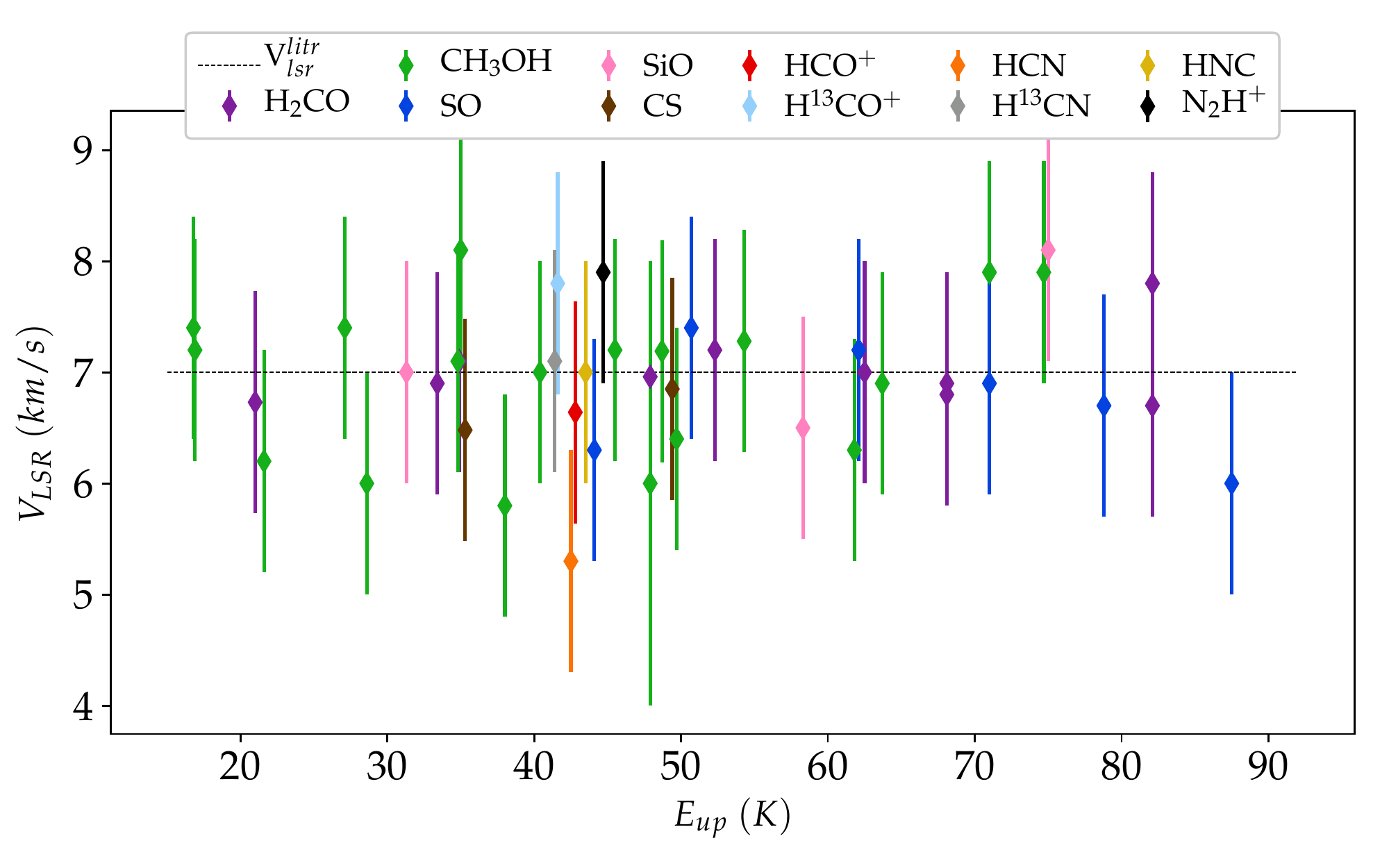}
            \includegraphics[width=0.48\textwidth]{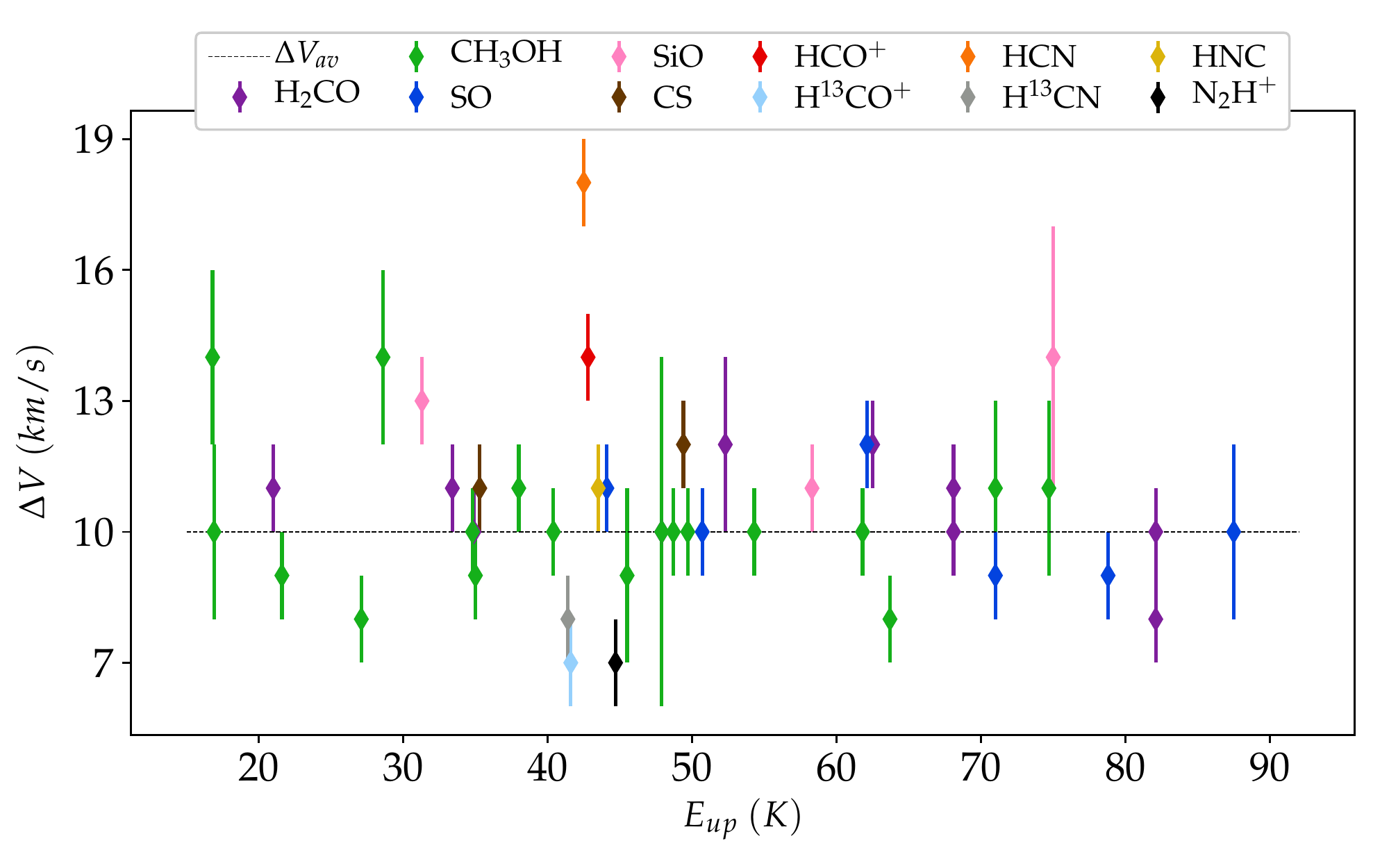}
            \caption{Scatter plots for LSR velocity, $V_{\rm LSR}$, (\textit{top}) and line widths, $\Delta V$, (\textit{bottom}) for detected lines of different species in W28\,F.}
            \label{scr:all}
        \end{figure}

        Figure\,\ref{scr:all} shows scatter plots of the LSR velocities {(upper panel)} and line widths {(lower panel)} of all detected lines as a function of the upper energies above ground state, $E_{\rm up}$, of the transitions (see Tab.\,\ref{tab:line}). There does not seem to be any particular dependence of either quantity on $E_{\rm up}$. The LSR velocities are scattered around $\sim\,7\,\si{\km\per\s}$. This is in agreement with W28's $V_{\rm LSR}$ values found in the literature \citep{ari,fukui2008,velazquez2002,reach2005}. Line widths are found in the range from 7 to 18$\, \si{\km\per\s}$, with most of the lines close to 10$\, \si{\km\per\s}$, indicating that they originate from the post-shock gas. Many plausible reasons may contribute to a large scatter in line widths. First, transitions with a low signal-to-noise ratio ($S/N$), e.g. \ce{CH3OH} ($1_{1}\,A^+ - 0_{0}\,A^+$) at 350905.100$\,$MHz and \ce{SiO} ($8-7$) at 347330.581$\,$MHz, have a large uncertainty in their fitted line width values. Second, species such as \ce{HCN} and \ce{HCO^+} are more abundant compared to others. Consequently, their emission lines are much stronger and have detectable broad wings. Their less abundant isotopologues like \ce{H^{13}CN} and \ce{H^{13}CO^+} on the other hand, emit lines that are much weaker and, hence, do not have detectable wings. Third, lines from HCN and \ce{HCO+} could also be broadened because of their high optical depths. Finally, species such as \ce{N_2H^+} may be tracing the ambient medium and, hence, demonstrate narrow line widths.
        \par In addition, we also checked that the hyperfine structure splitting for the (4--3) transitions of HCN, \ce{H^{13}CN}, \ce{N^2H^+}, and HNC and the relative intensities of the satellites are too small to have a measurable effect on the profiles of lines with a typical width of 10$\rm \,km\,s^{-1}$. Hereafter, we adopt $V_{\rm{LSR}}\,\sim\,7\,\si{\km\per\second}$ and $\Delta V\,\sim\,10\,\si{\km\per\second}$.

    \section{Modeling molecular line intensities : Physical conditions} \label{sec:phys}
      We used multiple lines from different species that we observed to probe the physical conditions in  W28$\,$F region. We used RADEX\footnote{\url{http://www.sron.rug.nl/~vdtak/radex/index.shtml}} \citep{radex} to model the observed line integrated intensity, which is a statistical equilibrium radiative transfer code made available for public use.
        It is a useful tool that provides constraints on physical conditions, namely density and kinetic temperature. RADEX computes the excitation of the molecular lines  under non-LTE conditions, provided their collisional rate coefficients are available. It {estimates} line intensities using the escape probability formulation. In order to calculate the escape probabilities, we use the Large Velocity Gradient (LVG) approximation (\citealt{1960mes..book.....S, 1977A&A....60..303S}). Under this approximation the emitted line photons can escape as long as the velocity shift resulting from the velocity gradient is larger than the local line thermal width, which is valid in C-type shocks (see e.g., \citealt{2008A&A...482..809G}). Other assumptions used in RADEX modeling include the source being an isothermal and homogeneous sphere filling the beam of the telescope. A detailed description of our RADEX analysis is described in the Appendix \ref{sec:appendix_formaldehyde}.

     {Since all the observations were done towards the direction of brightest emission from \ce{CH3OH} lines at 290$\,\si{\giga\hertz}$ near the OH maser regions, our data lacks the scope to allow us to determine the extent of the molecular emission. Looking at the extended nature of CO emissions mapped by  \cite{frail98} (see their Fig. 1a) and \cite{gusdorf} (see their Fig. 1), we may assume that the emission from other molecules is also extended in nature, which would correspond to a beam-filling factor of 1. However, this assumption may not be entirely true, in which case, a beam dilution factor needs to be assumed. Apart from the standard calibration error associated with APEX data, this unknown beam-filling factor is the largest source of uncertainty when scaling line intensities to account for differences in beamsize caused by different observing frequencies. In order to quantify how this impacts the results, all the analyses presented throughout this paper were carried out for two extreme cases: a point like source and an extended source covering the whole beam. The results obtained for a point-like source were not significantly different from those of the extended source when a 20\% uncertainty was assumed while scaling the line intensities for the latter. Therefore, all results presented in this paper correspond to the case of an extended source with a beam-filling factor of 1 along with a 20\% uncertainty added while scaling intensities to account for differences in beam size.} 

    \subsection{Formaldehyde \label{sec:h2co}}
         \begin{figure}
             \centering
             \includegraphics[width=0.4\textwidth]{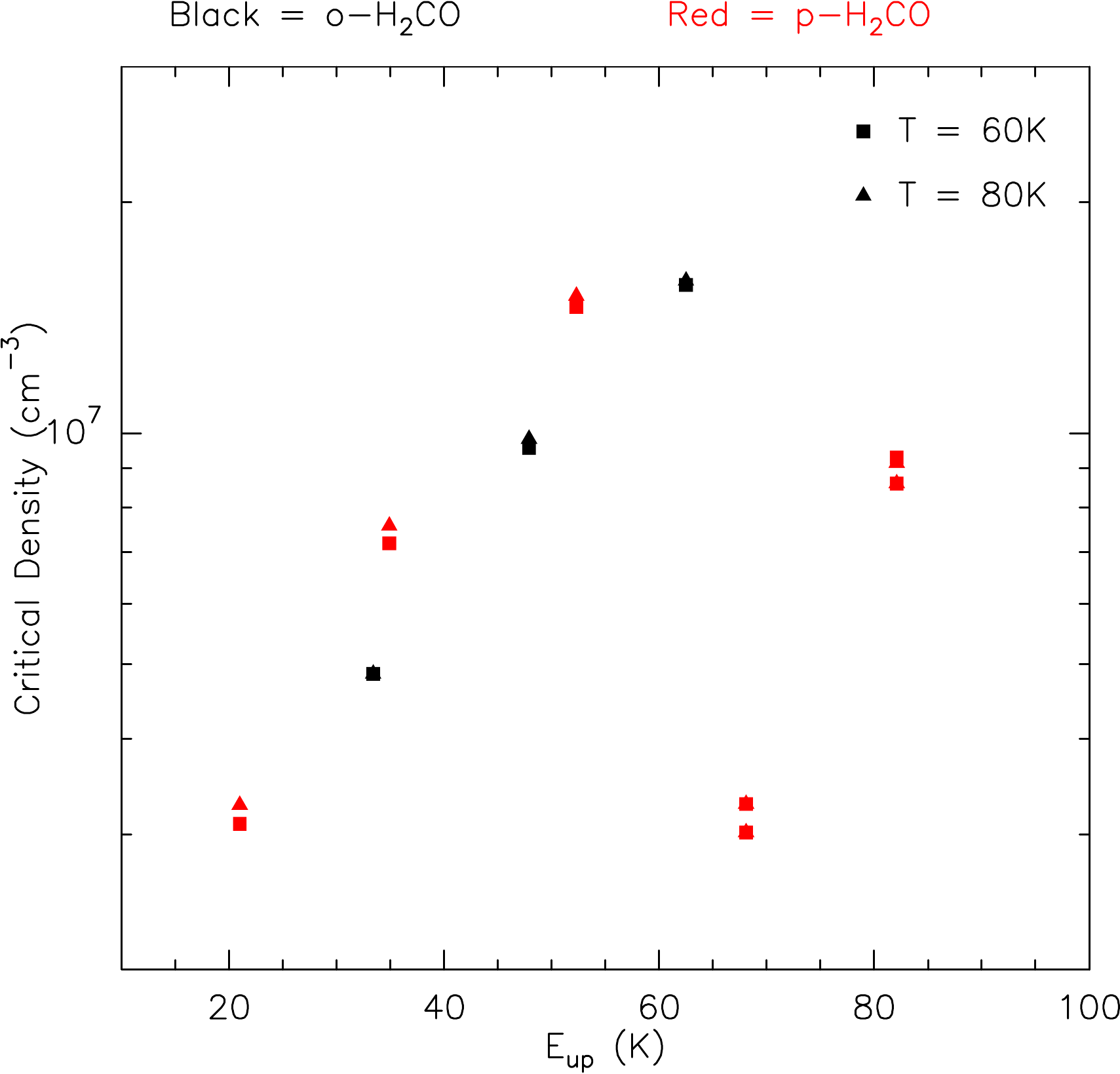}
             \caption{{Critical density versus upper level energy scatter plot of observed o-\ce{H2CO} (black symbols) and p-\ce{H2CO} (red symbols). Collisions only with \ce{H2} at 60$\,$K (square) and 80$\,$K (triangle) have been used to derive the critical densities.}}
             \label{fig:h2co_crit_density}
         \end{figure}

        \par Formaldehyde (H$_2$CO), a slightly asymmetric rotor molecule, is a ubiquitous molecule in  interstellar clouds and exhibits a large number of millimetre and sub-millimetre transitions. Previous observations of H$_2$CO in various molecular environments indicate that it is a reliable tracer of physical conditions of dense gas (see \citealt{man,gins2011,ao2013,gins2015,gins2016,tang2017b,tang2017a,tang2018a,tang2018b,man2019}).
        \begin{figure*}[ht]
            \centering
            \includegraphics[width=\textwidth]{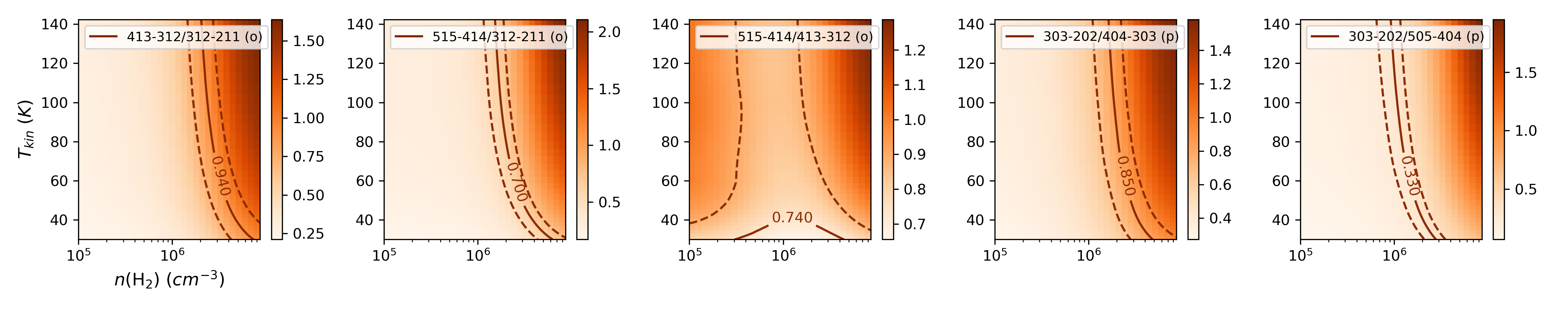}
            \includegraphics[width=\textwidth]{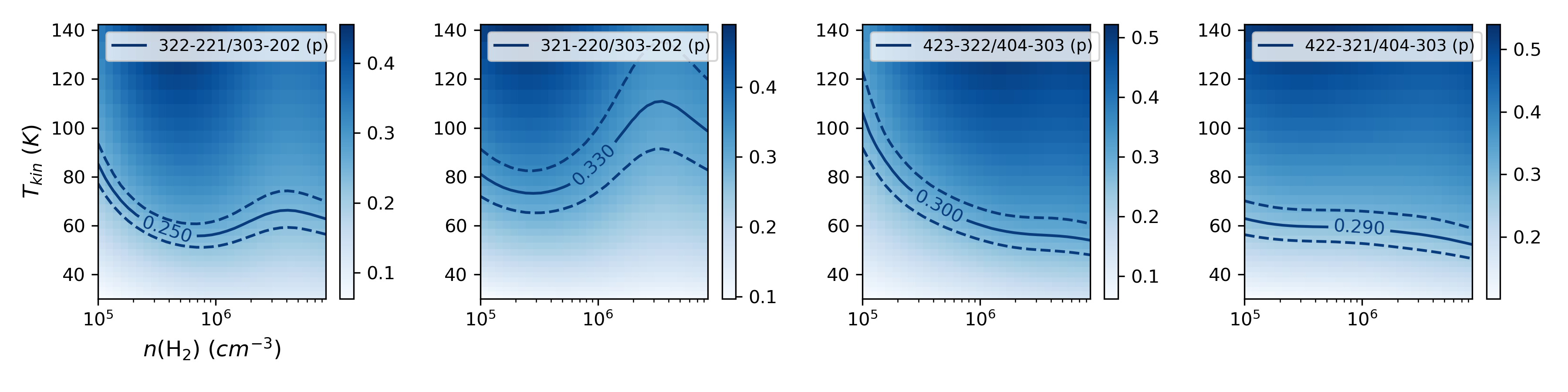}
            \caption{RADEX non-LTE modeling of the \ce{H2CO} line ratios to probe \ce{H2} spatial-density (\textit{top}) and kinetic temperature (\textit{bottom}). The background color shows the modeled ratios on the density and temperature grid. Solid lines are the contours of modeled ratios corresponding to the observed values and dashed lines correspond to the uncertainties. Adopted value of column density for these contours is $N = \num{e13}\,\si{\per\square\cm}$. The effects of optical depths were monitored during the modeling and they are smaller than 0.3 and hence do not effect the results significantly.}
            \label{fig:h2co_ratio_model}
        \end{figure*}
        \par The peak intensity of emission lines depend on the kinetic temperature ($T_{\rm kin}$), density of collision partners (e.g. $n_{\rm H_2}$) and column density ($N$(H$_2$CO)) of the molecule. However, when the density of the colliding partner {(e.g., H$_2$)} is higher than the critical density of the molecule, the collisions will be frequent enough to populate the energy levels of the molecule according to the Boltzmann distribution law. Consequently, the intensity of emission will only be a function of temperature and column density. The critical density of a transition from level $i$ to level $j$ is defined by
         \begin{equation} \label{eq:crit_density}
            n_{crit_{ij}} = A_{ij}/C_{ij} ,
        \end{equation}
        where $A_{ij}$ is the Einstein A coefficient and $C_{ij}$ is the collisional rate coefficient in $\si{\cubic\cm\per\second}$. 
        
         {Using the intensities of individual \ce{H2CO} lines to constrain the kinetic temperature and spatial densities 
         can be error-fraught due to large absolute calibration uncertainties and beam-filling factor. Thus, we opted to use line ratios instead. The relative populations of the \ce{H2CO} rotational energy levels in different $K_{\rm a}$ ladders are predominantly governed by collisions since dipole selection rules dictate that $\Delta K = 0$ for radiative excitation; hence, ratios of line fluxes involving different $K_{\rm a}$ ladders are good tracers of the kinetic temperature. On the other hand, line ratios involving lines emitted from within the same $K_{\rm a}$ ladders yield estimates of the spatial density of the gas \citep{man}.}
        We excluded the collisions with electrons and atomic hydrogen in our RADEX analysis because they are not expected to have a significant contribution in comparison to \ce{H2}. The abundance ratio of electrons ($x(e)= n(e)/n(\ce{H2}) \lesssim 10^{-5}$) should not be large given that the gas has undergone a $C$-type shock \citep{gusdorf}. Observations of W28 by \cite{velazquez2002} showed that overall molecular hydrogen is at least three times more abundant than H {\small{I}} in the SNR as a whole. Hence, neglecting collisions with electrons and hydrogen atoms is justified. Nevertheless, \cite{van} did not find a significant difference in results from including collisions from these species in their similar analysis of SNR IC\,443 either.
        
        Figure \ref{fig:h2co_crit_density} shows the critical densities of \ce{H2CO} lines observed in W28\,F. The values of the collisional coefficients have been taken from the Leiden Atomic and Molecular Database (LAMDA)\footnote{\url{http://home.strw.leidenuniv.nl/~moldata/}} \citep{sch}. The $C_{ij}$ value adopted for Fig.\,\ref{fig:h2co_crit_density} corresponds to $T_{\rm kin} = 60\,\si{\kelvin}$, with a caution that $C_{ij}$ varies as a function of the kinematic temperature. However, the lowest derived critical density for a range of $50<T_{\rm kin}<100\,\si{\kelvin}$ is $\sim\,\num{1e6}\,\si{\per\cubic\cm}$ and the dependence of $C_{ij}$ on kinetic temperature is likely insignificant for H$_{2}$CO (see Fig.\,\ref{fig:h2co_crit_density}). Since the critical densities of the observed lines can reach up to $\sim 2 \times 10^7 \, \si{\per\cubic\cm}$, the LTE condition is most likely not satisfied. Hence, non-LTE modeling of the emission is required to constrain the physical conditions existent in W28\,F {from formaldehyde emission lines}.

        \begin{figure*}[ht]
            \centering
            \includegraphics[width=\textwidth]{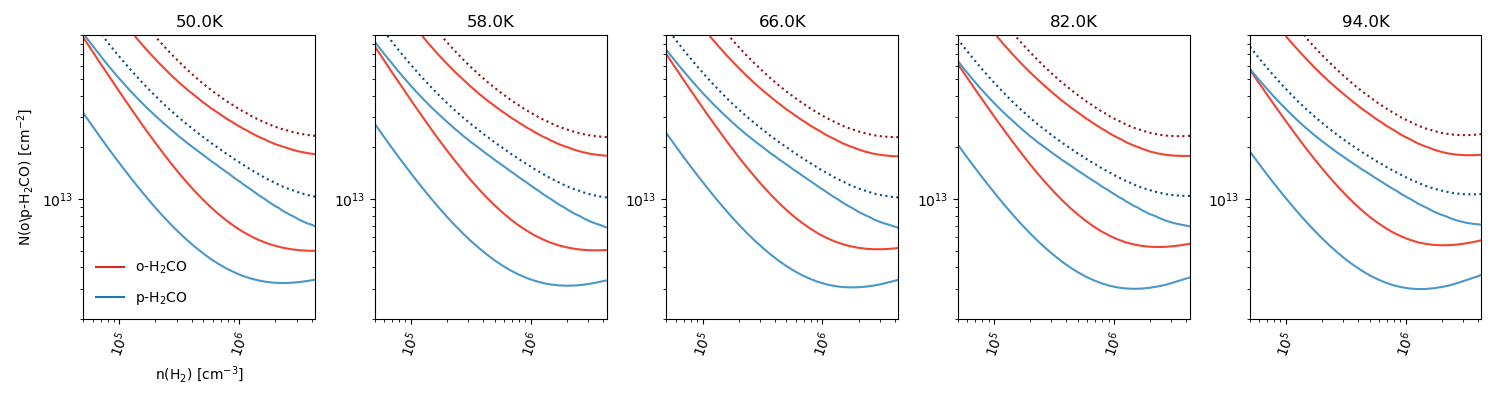}
            \caption{RADEX non-LTE modeling of the line intensity of o- and p-H$_{2}$CO to probe the column densities of both species. Each panel corresponds to a slice of constant $T_{\rm kin}$ from a 3D grid of $\Bar{\sigma}^2$ (see Appendix \ref{sec:appendix_formaldehyde}). Red contours are for o-\ce{H2CO}, while blue ones are for p-\ce{H2CO}. The solid and dotted contours are represented by 99.7$\%$ and 99.99$\%$ that the observed intensity inside the modeled contours (or $\Bar{\sigma}^2$ = 9 and 25), respectively.}
            \label{fig:h2co_int_model}
        \end{figure*}
                       
        \par Many pairs {of} H$_2$CO transitions can be measured simultaneously with the same receiver system and angular resolution. Being able to simultaneously observe multiple lines means the calibration uncertainties are largely reduced. Additionally, taking their flux ratio eliminates the uncertainties related to unknown beam-filling factors.
        {In this work, we used the lines ratios $3_{22}$--$2_{21}$/$3_{03}$--$2_{02}$, $3_{21}$--$2_{20}$/$3_{03}$--$2_{02}$, $4_{23}$--$3_{22}$/$4_{04}$--$3_{03}$, and $4_{22}$--$3_{21}$/$4_{04}$--$3_{03}$ of \ce{p-H2CO} to constrain the kinetic temperature. The density was constrained using the $4_{04}$--$3_{03}$/$3_{03}$--$2_{02}$ and $5_{05}$--$4_{04}$/$3_{03}$-$2_{02}$ line ratios of \ce{p-H2CO} as well as $5_{15}$--$4_{14}$/$3_{12}$--$2_{11}$ and $4_{13}$--$3_{12}$/$3_{12}$--$2_{11}$ of \ce{o-H2CO}.} The intensity ratios were modeled on a 2 dimensional grid of kinetic temperature ($T_{\rm kin}$) and \ce{H2} density ($n(\rm H_2)$) for a fixed of column density for both ortho- and para-\ce{H2CO}{, whose effect will be considered later}. The grid points were logarithmically distributed over the range of variables {as listed in Table \ref{tab:radex_grid}}. A value of $\Delta V$\,=\,10\,$\si{\km\per\second}$ was used throughout. The background temperature was assumed to be that of the CMB, $T_{\rm CMB}$\,=\,2.73\,$\si{\kelvin}$.
        
        \par Figure\,\ref{fig:h2co_ratio_model} shows the results of the RADEX modeling on the line ratios for $N(\ce{H2CO}) = 10^{13}\,\rm cm^{-2}$. {Firstly, the optical depth is less than 0.5 for all lines so that the extinction is rather small and could be neglected}. {Secondly, the gas number density is constrained as $n(\rm H_{2})\simeq 1-\num{5e6} \,\si{\per\cubic\cm}$ (see top panel). However, the line ratio of $5_{15}$--$4_{14}$/$4_{13}$--$3_{12}$ does not constrain the density very well.}  
        {We note that} the density constraints obtained from the observed line ratios for p-\ce{H2CO} are consistent with those obtained from o-\ce{H2CO}. So, both these lines are also most likely originating from a similar region. {Lastly, RADEX constrains the kinetic temperatures as $T_{\rm kin}\sim 50-80\,\si{\kelvin}$ with $n_{\rm H_{2}}>10^{6}\,\rm cm^{-3}$ (see bottom panel)}. However, the ratio of $3_{21}-2_{20}/3_{03}-2_{02}$ results in a higher constraint of $T_{\rm kin}$. {Such high values of kinetic temperature in addition to broad line widths also indicate that the emission is originating from the post-shock gas.}

        \par The column density was then varied from $10^{12}$ to $10^{14}\,\si{\per\square\cm}$, but no significant change in the results was seen. {For these estimates} of the kinetic temperature and the gas number density, we estimated the column density of $N(\rm{p}-\ce{H2CO}) \sim 0.3-\num{1e13}\,\si{\per\square\cm}$ and $N(\rm{o}-\ce{H2CO}) \sim 0.5-\num{2.5e13}\,\si{\per\square\cm}$ from 99.7\% confidence level contours in Fig. \ref{fig:h2co_int_model}.

    \subsection{Methanol  \label{sec:ch3oh}}

        \begin{figure}
            \centering
            \includegraphics[width=0.4\textwidth]{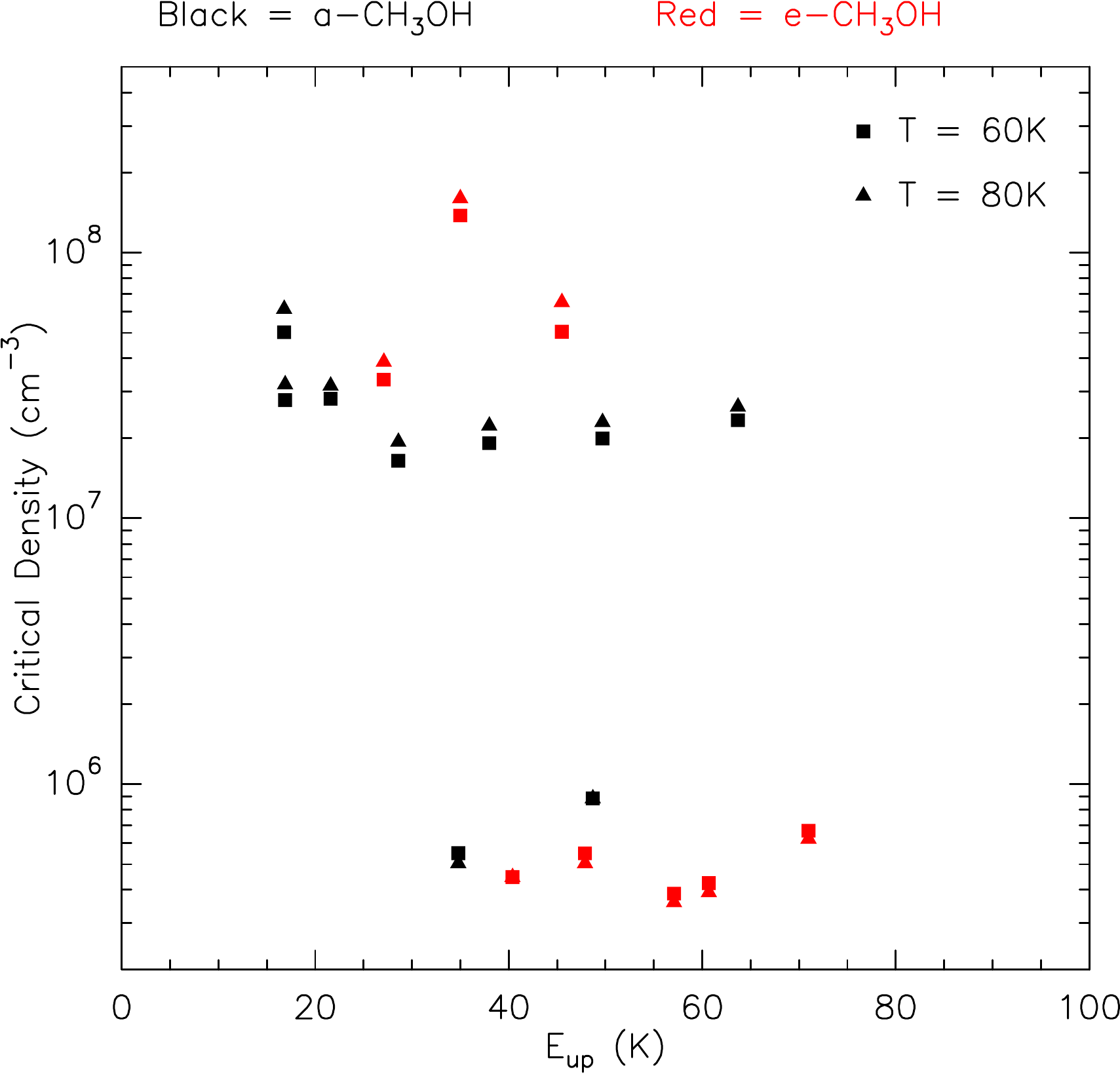}
            \caption{Critical density vs. upper level energy plot for the observed transitions of \ce{CH3OH}. {The collisional coefficient is achieved by only $H_{2}$ collider}.}
            \label{fig:ch3oh_crit_density}
        \end{figure}

        \par Methanol exists in two separate species, namely A-type CH$_3$OH and E-type CH$_3$OH. 
        All transitions from one to the other via collisions are forbidden, and de facto, as in the cases of ortho- and para-\ce{H2CO}, they are treated as independent species. Figure \ref{fig:ch3oh_crit_density} shows the critical densities of the observed A- and E-\ce{CH3OH} lines at $T_{\rm kin} = 60$ and $80\,\si{\kelvin}$, {considering only $H_{2}$ collisions}. Multiple transitions have a critical density of $\rm{1--5} \times 10^7\,\si{\per\cubic\cm}$. Thus, a non-LTE RADEX model is needed. The grid is listed in Table \ref{tab:radex_grid}. Similarly to the case of formaldehyde, {we used the ratios of lines to constrain on the kinetic temperatures and the gas number densities. However, the \ce{CH3OH} lines are far apart in frequency and were not observed simultaneously}. Therefore, the associated calibration and filling factor uncertainties cannot be eliminated.

\begin{figure}
            \centering
            \includegraphics[trim=0.4cm 0.2cm 0 0, clip,width=0.5\textwidth]{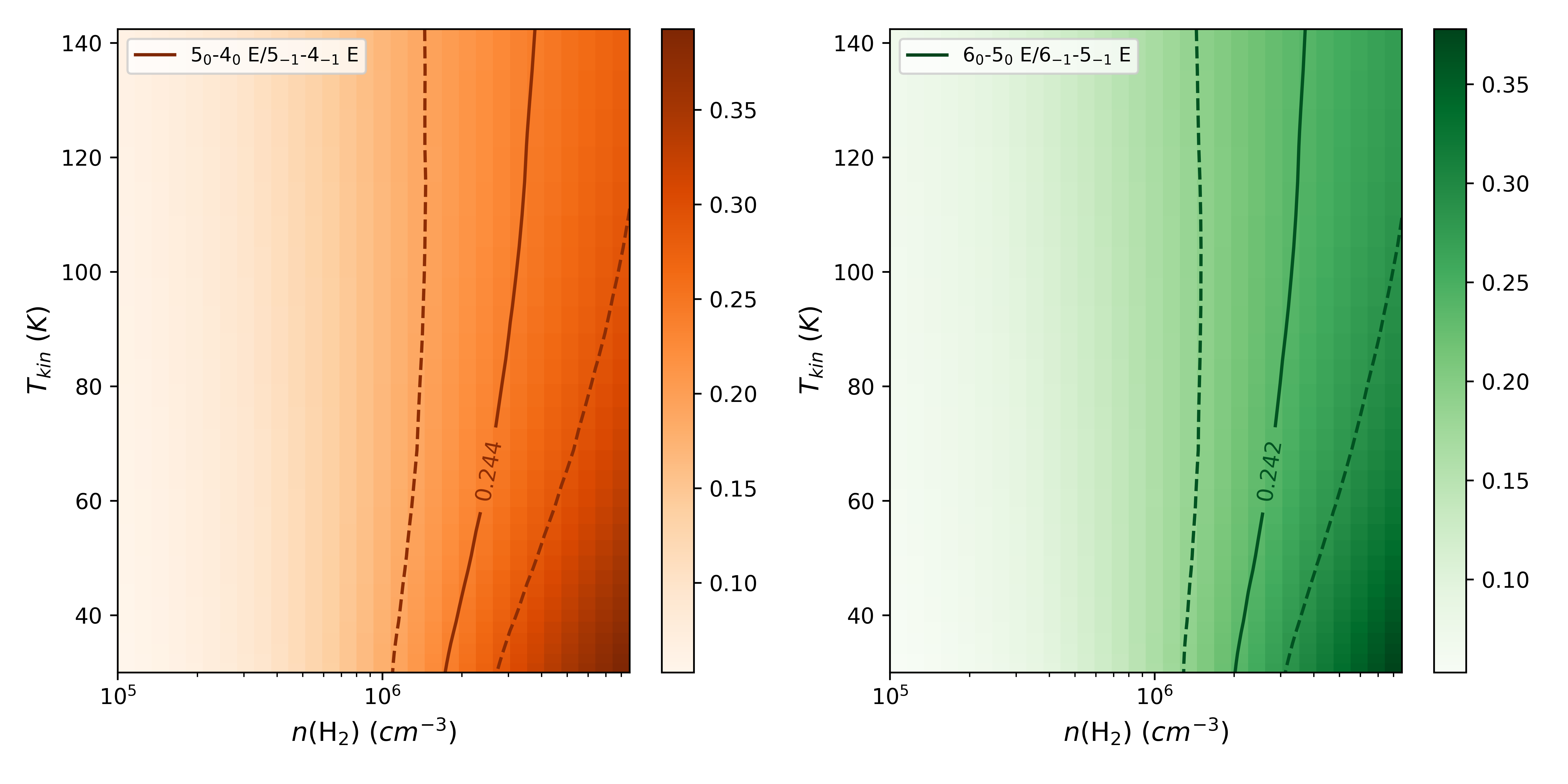}
            \caption{RADEX non-LTE modeling of the \ce{CH3OH} line ratios to probe \ce{H2} spatial-density. The solid lines are the observed values and dashed lines correspond to the uncertainties. {The adopted value of column density used to produce these contours is $N = \num{7e13}\,\si{\per\square\cm}$.}}
            \label{fig:ch3oh_ratio_model_density}
        \end{figure}

        \cite{Leu} demonstrated that the $5_k - 4_k$ and $7_k - 6_k$ line ratios of CH$_{3}$OH at 241$\,\si{\giga\hertz}$ and 338$\,\si{\giga\hertz}$ are sensitive enough to probe the dense gas, with density higher than $\num{e5}\,\si{\per\cubic\cm}$. Our observations did not cover the frequencies of the $7_k - 6_k$ series. Hence we {could only use the E-CH$_3$OH$\,5_{-1}$--$4_{-1}\,E/5_{0}$--$4_{0}\,E$ at 240$\,\si{\giga\hertz}$ along with the $6_{-1}$--$5_{-1}\,E/6_{0}$--$5_{0}\,E$ line ratio at 290$\,\si{\giga\hertz}$ instead}. Figure \ref{fig:ch3oh_ratio_model_density} shows the contours of observed line ratios modeled from RADEX with a fixed column density of $N=\num{7e13}\,\si{\per\square\cm}$ ({varying the column density value over the overlapping $3\sigma$ range from Fig. \ref{fig:ch3oh_int_model} did not change the results significantly}). {Both line ratios are} not sensitive to the kinetic temperature. However, if we adopt the kinetic temperature in the range of $50<T_{\rm kin}<80\,\si{\kelvin}$ from the formaldehyde analysis, {the density value is constrained to} $n(\rm H_2) = 2^{+4}_{-1} \times 10^{6}\,\si{\per\cubic\cm}$. This is in agreement with the result inferred from the \ce{H2CO}. Hence, the emission is probably originating from the same cloud. 
        
        {The observed lines intensities as described in Table. \ref{tab:line} were used to probe the CH$_{3}$OH column density. Figure \ref{fig:ch3oh_int_model} shows the contours constraining the observed line intensities of \ce{CH3OH} on an $N(\ce{CH3OH})$ versus $n_{\rm H_{2}}$ grid for different values of $T_{\rm kin}$. With $T_{\rm kin}=$60$\,\si{\kelvin}$ and $n_{\rm H_{2}} = 2 \times 10^{6}\,\si{\per\cubic\cm}$, the $3\sigma$ contours (99.7\% confidence level) of A- and E-\ce{CH3OH} indicate a column density of $4 \times 10^{13} \lesssim N(E-\ce{CH3OH}) \lesssim 1.5 \times 10^{14}\,\si{\per\square\cm}$ and $2 \times 10^{13} \lesssim N(A-\ce{CH3OH}) \lesssim 8 \times 10^{13}\,\si{\per\square\cm}$.}

        \begin{figure*}
            \centering
            \includegraphics[width=1\textwidth]{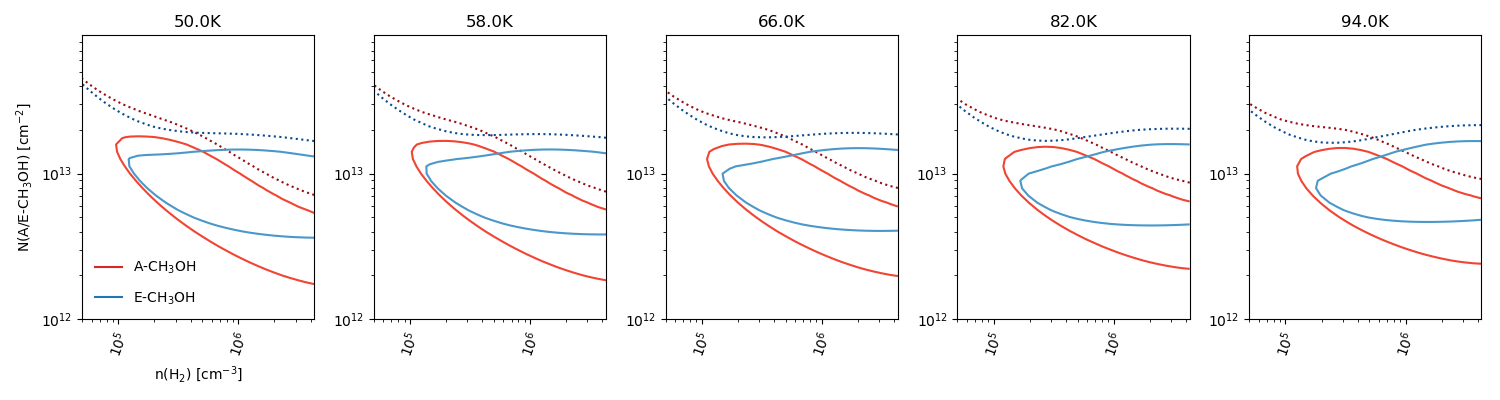}
            \caption{2D constant $T_{\rm kin}$ slices from a 3 dimensional grid of $\Bar{\sigma}^2$. Contours of constant $\Bar{\sigma}^2$ = 9 (solid); 25 (dotted) on a column density vs. \ce{H2} density grid for detected lines of A-\ce{CH3OH} ({red contours}) and E-\ce{CH3OH} ({blues contours}) are shown.}
            \label{fig:ch3oh_int_model}
        \end{figure*}

    \subsection{Other molecular lines  \label{sec:others}}

        We also attempted to use the other multiple transitions of SO, SiO, and CS towards W28\,F. {Similarly, we first calculated the critical densities with collision partner $H_{2}$ which results in the lowest critical density of $\num{2.4e6}\,\si{\per\cubic\cm}$ corresponding to the SO ($J=5_5-4_4$) line. Hence, these lines were also modeled using the non-LTE RADEX code}. However, the results appear to be independent of the choice of kinetic temperature over the range of modeled values (see Fig. \ref{fig:others_int_model}). Other species with one single emission line detected, for instance, \ce{HCO^{+}}, \ce{H^{13}CO^{+}} , \ce{HCN}, and \ce{HNC}, RADEX could not also provide explicit constraints on the kinematic temperature, gas density, and their column density. Therefore, we made the assumption that all line emissions are originating from the same physical conditions as \ce{H2CO}, which allowed us to derive the column density of these molecules for $T_{\rm kin}=58\,\si{\kelvin}$ and $n(\rm H_2) = \num{2e6}\,\si{\per\cubic\cm}$. The results are listed in Table \ref{tab:column_densities}.

        \begin{table}[h]
            \small
            \centering
            \caption{Column densities of observed species in W28$\,$F.} \label{tab:column_densities}
            \begin{tabular} {c c}
                \hline \hline \noalign{\smallskip}
                Molecule &  \parbox [c]{1.2cm}{\centering $N$ $(\si{\per\square\cm})$}\\
                \hline \noalign{\smallskip}
                \ce{p-H_2CO} & 0.3--1$\,\times \,10^{13}$ \\
                \ce{o-H_2CO} & 0.5--2.5$\,\times \,10^{13}$ \\
                \ce{E-CH_3OH} & 0.4--1.4$\,\times \,10^{14}$  \tablefootmark{a} \\
                \ce{A-CH_3OH} & 1--8$\,\times \,10^{13}$  \tablefootmark{a} \\
                \ce{SO} & 1--3$\,\times \,10^{13}$  \tablefootmark{a} \\
                \ce{SiO} & 1--4$\,\times \,10^{12}$  \tablefootmark{a} \\
                \ce{CS} & 0.5--2$\,\times \,10^{13}$  \tablefootmark{a} \\
                \ce{HCO^+} & 0.3--1$\,\times \,10^{13}$  \tablefootmark{a} \\
                \ce{H^{13}CO^+} & 0.8--3$\,\times \,10^{11}$  \tablefootmark{a} \\
                \ce{HCN} & 1--5$\,\times \,10^{13}$  \tablefootmark{a} \\
                \ce{HNC} & 0.8--2$\,\times \,10^{12}$  \tablefootmark{a} \\
                \hline
            \end{tabular}
            \tablefoot{\tablefoottext{a}{\footnotesize{For an assumed kinetic temperature $T_{\rm kin} = 58\,\si{\kelvin}$ and a spatial density $n$(H$_2$)\,=\,$\num{2e6}\,\si{\per\cubic\cm}$}}.}
        \end{table}

\section{Discussion \label{sec:discuss}}

    \subsection{Molecular composition}
        {In this section, we estimate the abundances of our detected molecules. In order to estimate molecular abundances, we need to know the \ce{H2} column density in W28\,F. We adopt the value constrained by \citet{gusdorf}, in which the authors used non-dissociative shock modeling to estimate $N(\rm CO)$ as $1.9\times 10^{18}\,\si{\per\square\cm}$. Assuming a ``standard'' ISM CO abundance of [$^{12}$CO]/[H$_2$]\,=\,$10^{-4}$ \citep[see][]{bolatto2013}, the H$_2$ column density is $N(\rm H_2)\,=\,1.9\times10^{22}\,\si{\per\square\cm}$. We note that, in the presence of dissociative shocks, both CO and \ce{H2} can be destroyed and, depending on their reformation pathway, this abundance value can change.}
    
        \begin{table}[h]
            \small
            \centering
            \caption{Abundances of observed species in W28\,F.} \label{tab:abundances}
            \begin{tabular} {c c c c}
                \hline \hline \noalign{\smallskip}
                Molecule & W28 F & IC\,443 \tablefootmark{a} & TMC-1 \tablefootmark{b}\\
                \hline \noalign{\smallskip}
                \ce{H_2CO} & 0.4--1.8$\,\times \,10^{-9}$ & 4 $\,\times \,10^{-9}$ & 2 $\,\times \,10^{-8}$ \\
                \ce{CH_3OH} & 0.3--1.1$\,\times \,10^{-8}$ & < 1$\,\times \,10^{-10}$ & 2 $\,\times \,10^{-9}$ \\
                \ce{SO} & 0.5--1.6$\,\times \,10^{-9}$ &  8 $\,\times \,10^{-9}$ & > 7 $\,\times \,10^{-9}$ \\
                \ce{SiO} & 0.5--2.1$\,\times \,10^{-10}$ & 8 $\,\times \,10^{-10}$ & < 2 $\,\times \,10^{-12}$ \\
                \ce{CS} & 0.3--1.0$\,\times \,10^{-9}$ & 3 $\,\times \,10^{-9}$ & > 5 $\,\times \,10^{-9}$ \\
                \ce{HCO^+} & 1.6--5.2$\,\times \,10^{-10}$ & 3 $\,\times \,10^{-9}$  & 8 $\,\times \,10^{-9}$ \\
                \ce{H^{13}CO^+} & 0.4--1.6$\,\times \,10^{-11}$ & < 1 $\,\times \,10^{-10}$ & 2 $\,\times \,10^{-10}$ \\
                \ce{HCN} & 0.5--2.6$\,\times \,10^{-9}$ & 8 $\,\times \,10^{-9}$ & 2 $\,\times \,10^{-8}$ \\
                \ce{HNC} & 0.4--1.0$\,\times \,10^{-10}$ & 1 $\,\times \,10^{-9}$ & 2 $\,\times \,10^{-8}$ \\
                \hline
            \end{tabular}
            \tablefoot{ \tablefoottext{a}{Towards clump G I \citep{van}.} \tablefoottext{b}{\citet{gratier2016,ohishi1992} and references in Table 8 of \citet{van} for \ce{SiO}, \ce{H^{13}CO^+}.}}
        \end{table}
    
    {The results are shown in Table \ref{tab:abundances}. Furthermore, we compare them to what was found for another well-known SNR, IC443 by \citealt{van} and the dark cloud TMC-1 (an unshocked region), reported in \citealt[and references therein]{gratier2016} for a same range of observed frequencies. The comparisons show that (1) the abundances of molecules in W28\,F derived from our analysis are mostly smaller to values derived for  SNR\,443, except for \ce{CH3OH}, and (2) the molecules in the shocked gas of W$\,$28 are less abundant than in the non-shocked gas of TMC-1, except SiO and \ce{CH3OH}. The abundances in IC 443 were determined using the same CO/\ce{H2} ratio of $10^{-4}$ whereas the abundances in TMC-1 were calculated with respect to an \ce{H2} column density, of $10^{22}\,\si{\per\square\cm}$ that was assumed to be constant. Naturally, the comparison of abundances between W28 F and IC 443 is more straightforward than the one between W28 F and TMC-1 . We note that both [$^{12}$CO]/[H$_2$] and N[\ce{H2}] contribute to the uncertainties of the abundance calculations.}

    \subsection{Physical properties in comparison with previous works}
        \cite{frail98} observed three \ce{H2CO} transitions $3_{03}$--$2_{02}$, $3_{22}$--$2_{21}$, and $5_{05}$--$4_{04}$ {towards W28 F} to determine both kinetic gas temperature and density. {Using the methods prescribed by \cite{man}, they used the intensity ratio $3_{03}$-$2_{02}$/$3_{22}$--$2_{21}$ of 3.25 to estimate the kinetic temperature $T_{\rm kin}$\,=\,80$\pm$10\,K.} From our data we get a value of $3.9(4)$ for the same ratio. Similarly, \citet{frail98} also used the $3_{03}$--$2_{02}$/$5_{05}$--$4_{04}$ ratio to determine the density, interpolating between the LVG models of \cite{man} using the kinetic temperature of 80\,K to obtain $n_{\ce{H2}}$\,=\,$\num{2e6}\,\si{\per\cubic\cm}$. Our results obtained form the non-LTE RADEX model are consistent with their results.

        An early theoretical work by \cite{Elitzur} showed that collisions of \ce{OH} with \ce{H2} can create a strong inversion of the 1720\,MHz line for a range of kinetic temperatures (25\,$\leq$\,$T_{\rm kin}$ $\leq$\,200\,K) and molecular gas densities ($\num{e3}\,\leq\,n_{\ce{H2}}\,\leq\,\num{e5}\,\si{\per\cubic\cm}$) that are typical of the conditions expected in cooling post shock clouds. \citet{pav2,pav1}, {and \citet{Lockett1999}} included the effects of far-infrared line overlap (due to thermal and turbulent motions), and used newly computed collisional cross sections between \ce{OH} and \ce{H2} to confirm the basic result of \citet{Elitzur} for a limited range of $T_{\rm kin}$\,=\,50--125\,K. Given the higher critical densities of the probed methanol lines, our observations probe material that has a higher density than what would generally be conducive to the OH 1720\,MHz line's inversion. We note that, in addition, the detected 36$\,$GHz \ce{CH3OH} emission \citep{philstrom14} as well as the thermal intensity peak of methanol in our observations are at a slight offset from the observed 1720\,MHz position. In addition, the OH maser and the submillimeter lines observed could have different distributions along the line of sight and probe different volumes for this reason.

        Studies of IC\,443G in \cite{van} suggested that there is low density component, with $n_{\ce{H2}}$\,$\sim\num{e5}\,\si{\per\cubic\cm}$ and $T_{\rm kin}$\,$\sim$\,80\,K, and a high density component, with $n_{\ce{H2}}$\,$\sim\,\num{3e6}\,\si{\per\cubic\cm}$, at a temperature of $\sim$200\,K. In our work, our lack of data for a sufficient number of higher energy \ce{H2CO} lines did not allow constraints on a two component gas. The results we obtain  for W28\,F indicate an emission region that has attributes from both of the IC\,443\,G components, with the kinetic temperature in W28\,F being similar to that of the low density component in IC\,443\,G, while the spatial density of the former is similar to the high density component of the latter.

\subsection{Ortho-para ratio of formaldehyde}

        {In Sect.\,\ref{sec:h2co} we found similar spatial densities from ortho- and para-H$_2$CO line ratios}. Therefore, assuming a spatial density of $n(\ce{H_2}) = 2^{+4}_{-1} \times 10^{6}\,\si{\per\cubic\cm}$ (see Fig.\,\ref{fig:ch3oh_ratio_model_density}), we could use Fig.\,\ref{fig:h2co_int_model} to constrain the ortho-para ratio. This was done by aligning the $3\sigma$ contours in Fig.\,\ref{fig:h2co_int_model}. {The resulting range} for the ortho-para ratio is $1.5 \lesssim o/p \lesssim 3$.

        \par{For a full discussion of how the ortho to para ratio of formaldehyde is affected by different formation mechanisms, refer to \cite{kah} and references therein. Here we try to give a brief overview of the major factors that influence this ratio. In the gas phase, \ce{H2CO} is formed mostly via a reaction of  \ce{CH_3^+} with O, but electronic recombination of \ce{H_3CO^+ may also make a contribution}. Hence, the ortho/para ratio of \ce{H2CO} depends on the ortho/para ratio of these precursors. The interconversion between ortho- and para-\ce{H2CO} can also affect their ortho/para ratio. But, as \citet{kah} pointed out, the lifetime of \ce{H2CO}, which is determined by destruction by molecular ions is shorter than the time required for the ortho/para interconversion. If only gas-phase processes are  
        important for  
        \ce{H2CO} formation, 
        irrespective of whether \ce{CH_3^+} or \ce{H_3CO^+} is the dominant precursor, \citet{kah} showed that the ortho/para ratio for formaldehyde should be $\sim$\,3\,--\,5 at $10\,\si{\kelvin}$ and $\sim$\,3 at $70\,\si{\kelvin}$. However, atoms and molecules can get adsorbed onto grains and form \ce{H2CO} on their surface where their ortho/para ratio is likely to be thermalized at the grain temperature (i.e. the ratio corresponds to the Boltzmann distribution of ortho and para levels). Alternatively, \ce{H2CO} formed in the gas phase can get frozen out  on the surface of grains and undergo ortho to para conversion if the grains contain paramagnetic species or magnetic nuclei, thereby lowering their ortho-para ratio \citep{dickens-1999}. An ortho/para ratio $<$\,3, which we find for W28\,F, is therefore indicative of grain processing; i.e., either \ce{H2CO} was formed on grain surfaces followed by desorption from their surface, or, was adsorbed on grain surfaces after gas phase formation, processed there and then released back into a gas phase. Multiple mechanisms may be responsible for desorption of formaldehyde from grain surfaces, for example, thermal evaporation following Boltzmann's law, sporadic heating by cosmic rays \citep{hasegawa1993}, photodissociation by external radiation field, and via exothermic surface reactions \citep{duley1993,garrod2007}.}
        
        \par \citet{mangum1993} studied multiple (ortho and para) \ce{H2CO} in a sample of star forming regions and use their data to determine kinetic temperatures and ortho-para ratios. Generally, they found relatively high values for the temperatures $> 70$K and  typical ortho-para ratios less than 3. In line with our findings, they concluded that ``dust grains play an important role in \ce{H2CO} chemistry.''

\section{Conclusions \label{sec:concl}}
    \par We carried out a multi-molecular sub-millimeter wavelength line survey of the interaction zone of a molecular cloud with the supernova remnant W28. The APEX 12\,m telescope was used to observe one of the locations (F) of the 1720\,MHz OH maser emissions in a number of frequency ranges of the 230 and 345\,GHz atmospheric windows. Emission from multiple molecular species, {among which \ce{H2CO, CH3OH, SO, SiO, CS, HCO+} were detected}.
    The lines' centroid velocities  and widths indicate that they all are originating from the same gas at $V_{\rm LSR}\sim\,7\,\si{\km\per\second}$ and $\Delta V \sim \, 10\,\si{\km\per\second}$. Non-LTE radiative transfer modeling  of the lines with RADEX was used to determine the physical conditions of the cloud.
    For species with multiple line this modeling resulted in determinations of the kinetic temperature and \ce{H2} density: formaldehyde lines  
    were used to determine the kinetic temperature ($T_{\rm kin}\,=\,50-80\,\si{\kelvin}$) {and densities ($n$(H$_2$)$\,=1-{5}\times10^6\,\si{\per\cubic\cm}$)}. 
    For the same reason, methanol line ratios from the same observing windows were used to constrain the \ce{H2} density of the emitting cloud ($n$(H$_2$)$\,=\,2^{+3}_{-1}\times10^6\,\si{\per\cubic\cm}$). We constrained the ortho-to-para ratio of formaldehyde to the range of $1.5 \lesssim o/p \lesssim 3$. This indicates formation of formaldehyde {was most likely formed} on the surface of dust grains instead of the gaseous phase. Absolute line intensities were modeled using RADEX to constrain the column densities of the emitting species by {adopting the previously derived values for} kinetic temperature and \ce{H2} density. Using literature values for the CO column density, we were then able to put constraints on the abundances of various molecules.

\begin{acknowledgements}
    {We are indebted to the anonymous referee for invaluable input and many suggestions that resulted in a very significant improvement of this paper.}
    We thank the staff of the APEX telescope for their assistance in observations.
    This work acknowledges support by The Collaborative Research Council 956, subproject A6,
    funded by the Deutsche Forschungsgemeinschaft (DFG).
    X.T. acknowledges support by The Heaven Lake Hundred-Talent Program of Xinjiang
    Uygur Autonomous Region of China and The National Natural Science Foundation of China
    under grant 11903070 and 11433008.
    This research has used NASA's Astrophysical Data System (ADS).
\end{acknowledgements}

\bibliographystyle{aa} 
\bibliography{bibfile} 

\begin{appendix}

\onecolumn

\section{Observed spectra towards W28\,F}

    \begin{figure*}[!h]
      \centering
      \includegraphics[width=0.95\textwidth]{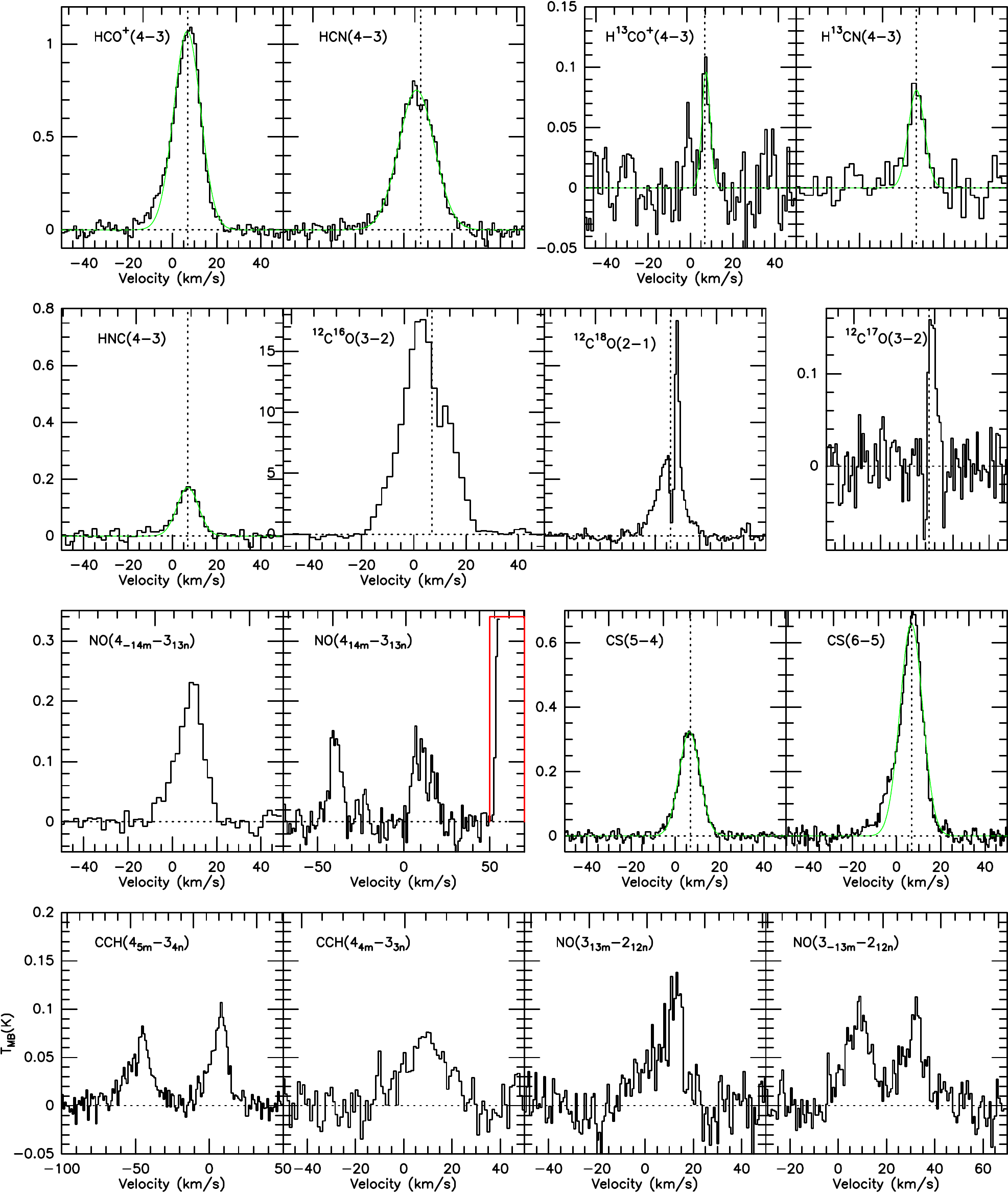}
      \caption{continued from Fig.\ref{fig:lines-1}. The spectral resolution lies between 0.8\,$\si{\km\per\second}$ and 1.2\,$\si{\km\per\second}$.}\label{spec}
    \end{figure*}
    
    \begin{figure*}[!htb]
      \ContinuedFloat
      \centering
      \includegraphics[width=0.98\textwidth]{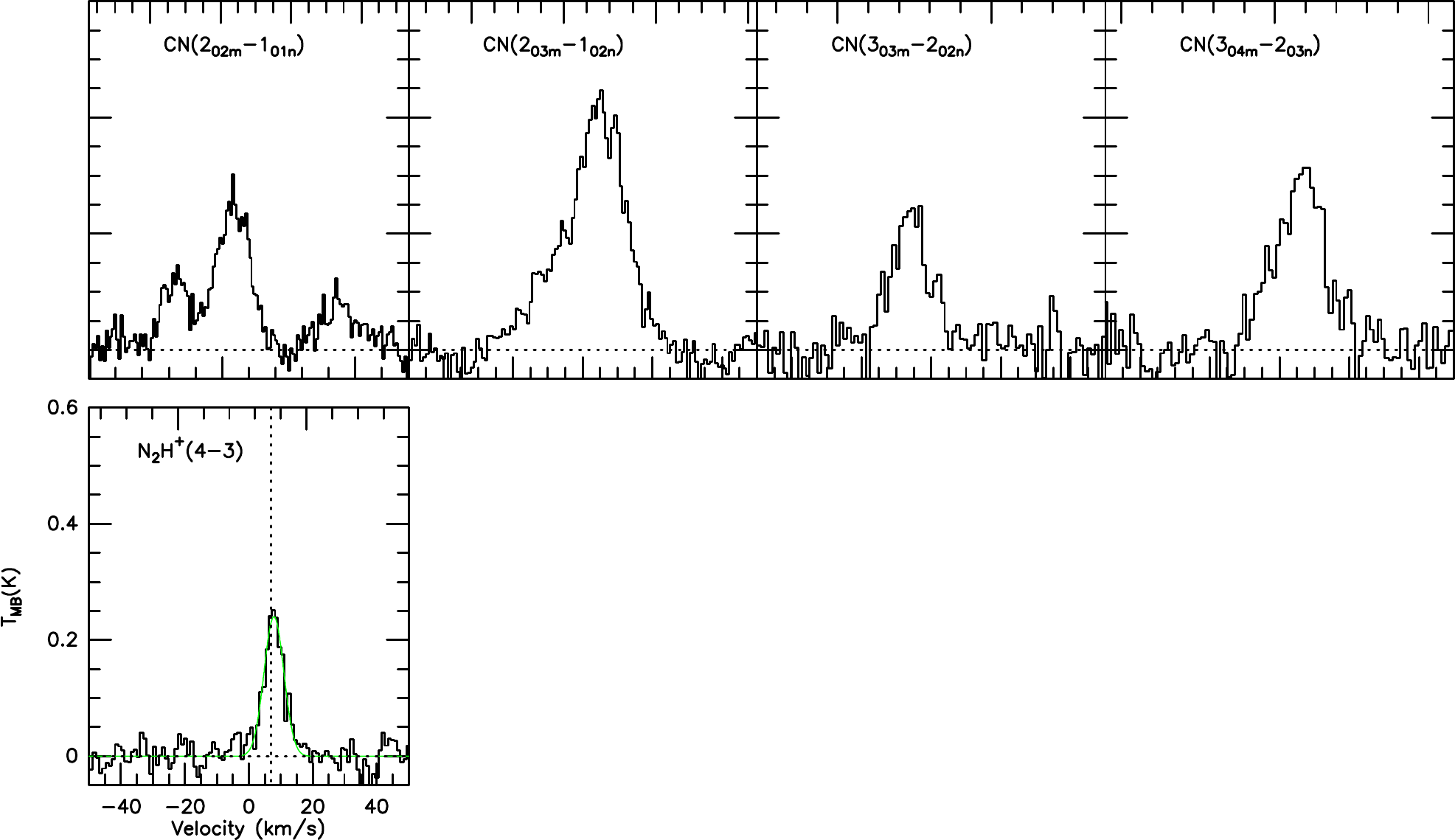}
    \caption{continued.}
    \end{figure*}

\clearpage

\section{Parameters of observed spectra towards W28\,F}
    \begin{table}
        \centering
        \caption{Table. \ref{tab:line} continued.} \label{tab:lines_cont}

    \begin{tabular}{l l c c c c c c}
        \hline \noalign{\smallskip}

        Species & Transition & \parbox[c]{2cm}{\centering Frequency\\$(\si{\MHz})$} & \parbox[c]{2cm}{\centering $E_{\rm up}$\\$(\si{\kelvin})$} & \parbox[c]{2cm} {\centering $\int$$T_{\rm MB}$ d$v$\\ $(\si{\K\km\per\s})$} & \parbox[c]{2cm}{\centering $V_{\rm LSR}$\\$(\si{\km\per\s}$)} & \parbox[c]{2cm}{\centering $\Delta V$\\$(\si{\km\per\s}$)} & \parbox[c]{2cm}{\centering $T_{\rm MB}$\\$(\si{\kelvin}$)} \\

        \hline \noalign{\smallskip}

        \ce{CN} & $2_{0,2,1} - 1_{0,1,2}$ & 226616.571 & 16.3 & - & - & - & $<0.06$ \tablefootmark{b}\\ 
        & $2_{0,2,2} - 1_{0,1,2}$ & 226632.190 & 16.3 & - & - & - & $<0.11$ \tablefootmark{d} \\ 
        & $2_{0,2,3} - 1_{0,1,2}$ & 226659.558 & 16.3 & - & - & - & $<0.13$ \tablefootmark{d} \\ 
        & $2_{0,2,1} - 1_{0,1,1}$ & 226663.693 & 16.3 & - & - & - & $<0.13$ \tablefootmark{d} \\ 
        & $2_{0,2,2} - 1_{0,1,1}$ & 226679.311 & 16.3 & - & - & - & $<0.13$ \tablefootmark{d} \\ 
        & $2_{0,3,3} - 1_{0,2,2}$ & 226874.191 & 16.3 & - & - & - & $<0.14$ \tablefootmark{d} \\ 
        & $2_{0,3,4} - 1_{0,2,3}$ & 226874.781 & 16.3 & - & - & - & $<0.14$ \tablefootmark{d} \\ 
        & $2_{0,3,2} - 1_{0,2,1}$ & 226875.896 & 16.3 & - & - & - & $<0.14$ \tablefootmark{d} \\ 
        & $2_{0,3,2} - 1_{0,2,2}$ & 226887.420 & 16.3 & - & - & - & $<0.06$ \tablefootmark{b} \\ 
        & $2_{0,3,3} - 1_{0,2,3}$ & 226892.128 & 16.3 & - & - & - & $<0.06$ \tablefootmark{b} \\ 
        & $3_{0,3,3} - 2_{0,2,3}$ & 340008.126 & 32.6 & - & - & - & $<0.10$ \tablefootmark{b} \\ 
        & $3_{0,3,2} - 2_{0,2,2}$ & 340019.626 & 32.6 & - & - & - & $<0.10$ \tablefootmark{b} \\ 
        & $3_{0,3,4} - 2_{0,2,3}$ & 340031.549 & 32.6 & - & - & - & 0.07(2) \tablefootmark{a} \\ 
        & $3_{0,3,2} - 2_{0,2,1}$ & 340035.408 & 32.6 & - & - & - & 0.07(2) \tablefootmark{a} \\ 
        & $3_{0,3,3} - 2_{0,2,2}$ & 340035.408 & 32.6 & - & - & - & 0.07(2) \tablefootmark{a} \\ 
        & $3_{0,4,4} - 2_{0,3,3}$ & 340247.770 & 32.7 & - & - & - & 0.10(2) \tablefootmark{a} \\ 
        & $3_{0,4,5} - 2_{0,3,4}$ & 340247.770 & 32.7 & - & - & - & 0.10(2) \tablefootmark{a} \\ 
        & $3_{0,4,3} - 2_{0,3,2}$ & 340248.544 & 32.7 & - & - & - & 0.10(2) \tablefootmark{a} \\ 
        & $3_{0,4,3} - 2_{0,3,3}$ & 340261.773 & 32.7 & - & - & - & $<0.10$ \tablefootmark{b} \\ 
        & $3_{0,4,4} - 2_{0,3,4}$ & 340264.949 & 32.7 & - & - & - & $<0.10$ \tablefootmark{b} \\ 
        \ce{CCH} & $4_{5,4} - 3_{4,4}$ & 349312.832 & 41.9 & - & - & - & $<0.03$ \tablefootmark{b} \\ 
        & $4_{5,5} - 3_{4,4}$ & 349337.706 & 41.9 & 0.68(7) & - & - & 0.04(1) \tablefootmark{a} \\ 
        & $4_{5,4} - 3_{4,3}$ & 349338.988 & 41.9 & 0.68(7) & - & - & 0.04(1) \tablefootmark{a} \\ 
        & $4_{4,4} - 3_{3,3}$ & 349399.276 & 41.9 & 0.66(7) & - & - & 0.03(1) \tablefootmark{a} \\ 
        & $4_{4,3} - 3_{3,2}$ & 349400.671 & 41.9 & 0.66(7) & - & - & 0.03(1) \tablefootmark{a} \\ 
        & $4_{4,3} - 3_{3,3}$ & 349414.643 & 41.9 & - & - & - & $<0.03$ \tablefootmark{b} \\ 
        \ce{NO} & $3_{1,3,4} - 2_{-1,2,3}$ & 250436.848 & 19.2 & - & - & - & $<0.06$ \tablefootmark{b} \\ 
        & $3_{1,3,3} - 2_{-1,2,2}$ & 250440.659 & 19.2 & - & - & - & $<0.06$ \tablefootmark{b} \\ 
        & $3_{1,3,2} - 2_{-1,2,1}$ & 250448.530 & 19.2 & - & - & - & $<0.06$ \tablefootmark{b} \\ 
        & $3_{-1,3,4} - 2_{1,2,3}$ & 250796.436 & 19.3 & - & - & - & $<0.10$ \tablefootmark{d} \\ 
        & $3_{-1,3,3} - 2_{1,2,2}$ & 250815.594 & 19.3 & - & - & - & $<0.06$ \tablefootmark{b} \\ 
        & $3_{-1,3,2} - 2_{1,2,1}$ & 250816.954 & 19.3 & - & - & - & $<0.06$ \tablefootmark{b} \\ 
        & $4_{-1,4,5} - 3_{1,3,4}$ & 350689.494 & 36.1 & - & - & - & 0.05(1) \tablefootmark{c,a} \\ 
        & $4_{-1,4,4} - 3_{1,3,3}$ & 350690.766 & 36.1 & - & - & - & 0.05(1) \tablefootmark{c,a} \\ 
        & $4_{-1,4,3} - 3_{1,3,2}$ & 350694.772 & 36.1 & - & - & - & 0.05(1) \tablefootmark{c,a} \\ 
        & $4_{1,4,5} - 3_{-1,3,4}$ & 351043.524 & 36.1 & - & - & - & 0.05(1) \tablefootmark{a} \\ 
        & $4_{1,4,4} - 3_{-1,3,3}$ & 351051.705 & 36.1 & - & - & - & 0.05(1) \tablefootmark{a} \\ 
        & $4_{1,4,3} - 3_{-1,3,2}$ & 351051.705 & 36.1 & - & - & - & 0.05(1) \tablefootmark{a} \\ 
        \ce{CS} & $5 - 4$ & 244935.557 & 35.3 & 3.57(5) & 6.5(1) & 11.0(2) & 0.31(1) \\ 
        & $6 - 5$ & 293912.086 & 49.4 & 8.34(8) & 6.8(1) & 11.6(1) & 0.67(2) \\ 
        \ce{HCO^+} & $4 - 3$ & 356734.223 & 42.8 & 16.6(1) & 6.6(1) & 14.5(2) & 1.07(3) \\ 
        \ce{H^{13}CO^+} & $4 - 3$ & 346998.344 & 41.6 & 0.46(5) & 7.8(3) & 7(1) & 0.06(1) \\ 
        \ce{HCN} & $4 - 3$ & 354505.477 & 42.5 & 14.4(3) & 5.3(2) & 17.8(4) & 0.76(5) \\ 
        \ce{H^{13}CN} & $4 - 3$ & 345339.769 & 41.4 & 0.74(10) & 7.1(5) & 8(1) & 0.08(2) \\ 
        \ce{HNC} & $4 - 3$ & 362630.303 & 43.5 & 2.01(10) & 7.0(3) & 11.0(6) & 0.17(2) \\ 
        \ce{N_2H^+} & $4 -3$ & 372672.481 & 44.7 & 1.80(10) & 7.9(2) & 7.0(5) & 0.24(3) \\ 
        \ce{CO} & $3 - 2$ & 345795.990 & 33.2 & - & - & - & 17.70(2) \tablefootmark{e} \\ 
        \ce{C^{18}O} & $2 - 1$ & 219560.354 & 15.8 & - & - & - & 0.59(1) \tablefootmark{e} \\ 
        \ce{C^{17}O} & $3 - 2$ & 337061.130 & 32.4 & - & - & - & 0.16(2) \tablefootmark{e}\\ 
        \hline
    \end{tabular}
    \tablefoot{The $T_{\rm MB}$ values are the fit results and not the absolute peak observed values. In cases where no Gaussian fitting could be done, the {observed temperature peaks} have been shown. \\ \tablefoottext{a}{These lines were blended with same $E_{\rm up}$ and $A_{\rm ij}$ values. They were fitted with one Gaussian and the integrated intensities were obtained by dividing the result of the fit by the number of blended lines assuming equal contribution from each. No attempt was made to derive LSR velocities or line widths.} \\
    \tablefoottext{b}{$3\sigma$ detection threshold applied for non-detected lines.} \\
    \tablefoottext{c}{{\ce{NO} lines were blended with a \ce{CH3OH} line and hence only an estimate of the upper limit on the peak temperature could be established. For this, we assumed that the NO lines at $\sim$351050 and $\sim$350690\,$\si{\mega\hertz}$ have similar emission strengths because of their identical $E_{\rm up}$ and very similar $A_{\rm ij}$ values.}} \\
    \tablefoottext{d}{{Multiple detected lines very closely spaced in frequency resulting a non-gaussian profile.} No Gaussian fitting was done. Emission upper limits were determined from peak $T_{\rm MB}$ values.} \\
    \tablefoottext{e}{CO line and its isotopes showed self absorption. {Multiple Gaussian line fitting was not performed in this study.} Hence, no Gaussian fitting was done to these lines.}
    }
\end{table}

\twocolumn

\section{Absolute intensity modeling with RADEX \label{sec:appendix_formaldehyde}}
    \par {All the integrated intensities were modeled on a 3D grid of kinetic temperature ($T_{\rm kin}$), \ce{H2} density ($n(\rm H_2)$) and column density ($N$). Prior to modeling, the intensities were scaled to account for the differences in beamsize owing to different frequencies. A 20\% uncertainty was added to the scaled intensities to account for the unknown beam-filling factor. The grid points were logarithmically distributed over the range of variables. A value of $\Delta V$\,=\,10\,$\si{\km\per\second}$ was used throughout. The background temperature was assumed to be that of the CMB, $T_{\rm CMB}$\,=\,2.73\,$\si{\kelvin}$. \ce{H2} was considered as the only collision partner (see Sect. \ref{sec:phys}).} 
     
    \par {The output of RADEX was then compared to the observed intensities via an averaged sum of squared deviations at each grid point calculated by,
    \begin{equation} \label{eq:sigma}
        \Bar{\sigma^2} = \left.{\sum_i^{n_{lines}} \frac{(W_{\rm obs_i}-W_{\rm mod_i})^2}{\sigma^2_{\rm W_{obs_i}}}} \middle/ {n_{lines}} \right. ,
        \end{equation}
    where $n_{lines}$ is equal to the total number of observed transitions. The contours of constant $\Bar{\sigma^2}$ were then plotted on the density versus column density grid for a constant temperature. Also, $\Bar{\sigma^2} = 9$ stands for $\sim$99.7\%, making it likely that the intensity observed is due to a grid point inside this contour. The $\Bar{\sigma^2} = 25$ contour means that $\sim$99.99\% likely that the intensity observed is due to a grid point inside this contour.}
     
     \begin{table*}
        \small
        \centering
        \caption{RADEX grid.} \label{tab:radex_grid}
        \begin{tabular} {c c c c c c}
            \hline \hline \noalign{\smallskip}
            \multicolumn{6}{c}{$T_{\rm kin}$\, vs.\, $n(\rm H_{2})^{a}$}\\
            \hline \noalign{\smallskip}
            T$_{\rm kin}$ & \multicolumn{5}{c}{30 - 150\, K} \\
            n(H$_{2}$) & \multicolumn{5}{c}{$10^5$ - $10^7$\, \rm cm$^{-3}$} \\
            \hline \noalign{\smallskip}
            \multicolumn{6}{c}{N\, vs.\, $n(\rm H_{2})$}\\
            \hline \noalign{\smallskip}
            {} & \ce{H2CO} & \ce{CH3OH} & \ce{SO} & \ce{SiO} & \ce{CS} \\
            N {[$\si{\per\square\cm}$]}  & $10^{12}-10^{14}$ & $10^{13}-10^{15}$ & $7\times 10^{12} - 10^{15}$ & $5\times 10^{11}-10^{14}$ & $3\times 10^{12} - 10^{15}$ \\
            $n(H_{2})$ {[$\si{\per\cubic\cm}$]} & $5\times 10^{4} - 5\times 10^{6}$ & $5\times 10^{4} - 5\times 10^{6}$ & $5\times 10^{4} - 10^{7}$ & $5\times 10^{4} - 10^{7}$ & $10^{5} - 10^{8}$ \\
            $T_{\rm kin}$ {[$\si{\kelvin}$]} & $50-100$ & $50-100$ & $50-100$ & $50-100$ & $50-100$ \\
            \hline
        \end{tabular}
        \tablefoot{\tablefoottext{a}{\footnotesize{N is fixed but its effect are shown in the N vs $n(H_{2})$}}.}
    \end{table*}
    
\section{RADEX modeling results for other molecules \label{sec:appendix_other_molecules}}
\begin{figure*}
        \centering
        \includegraphics[width=\textwidth]{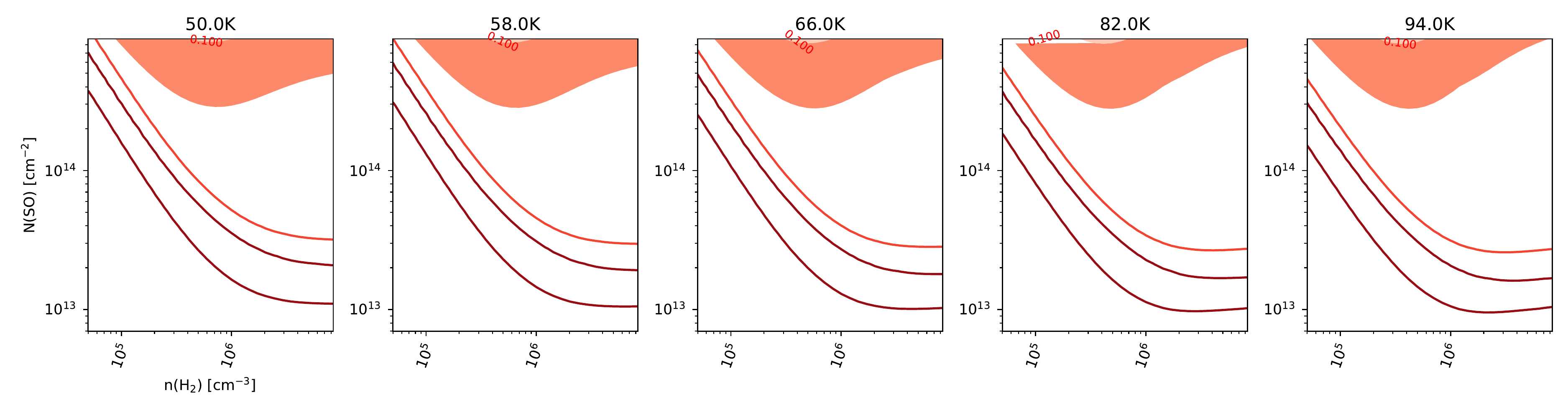} \\
        \includegraphics[width = \textwidth]{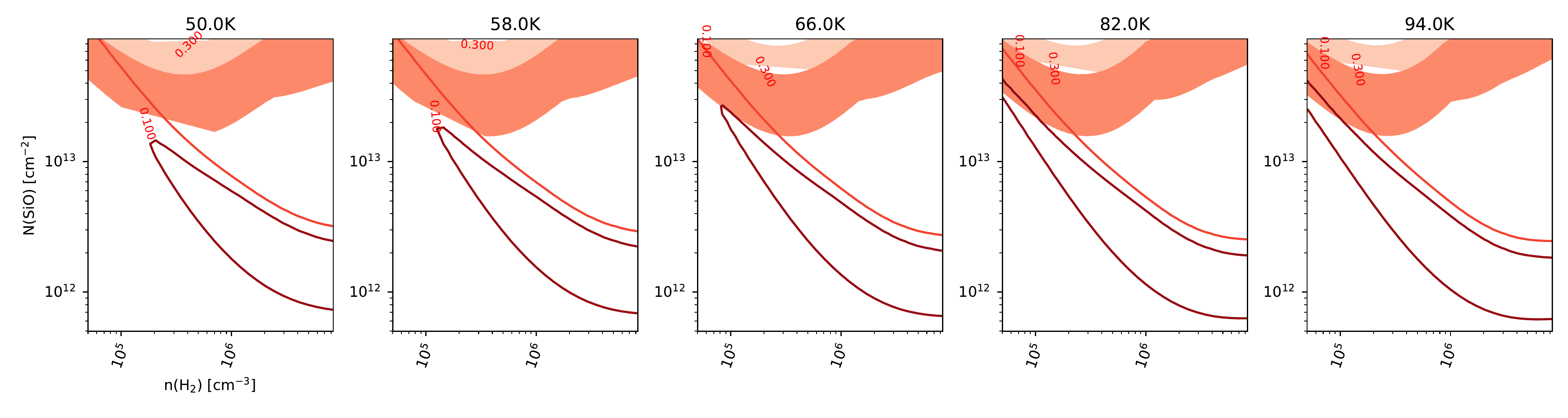} \\
        \includegraphics[width = \textwidth]{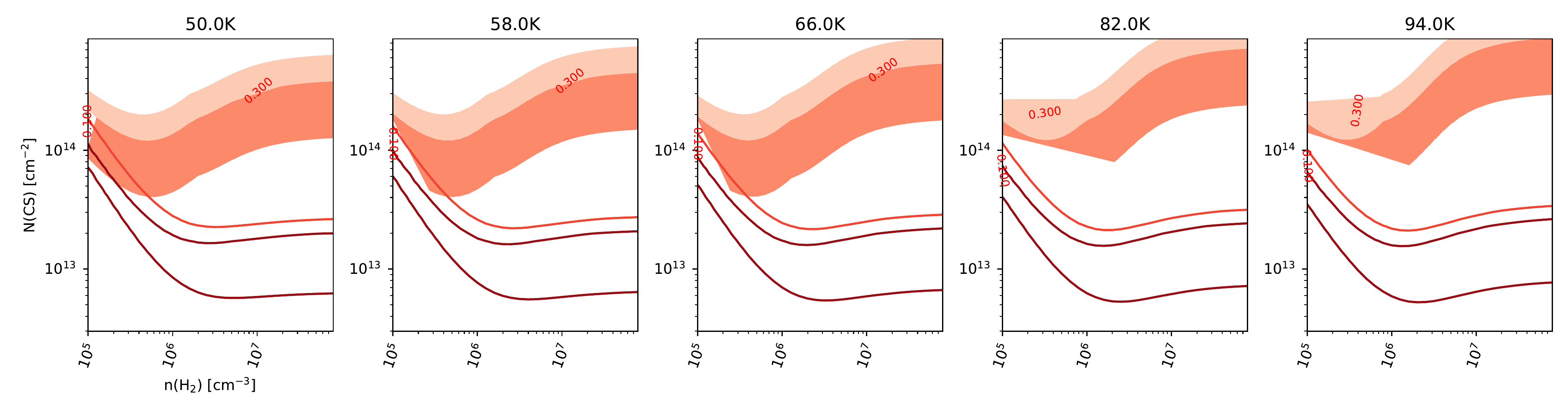}
        \caption{2D constant $T_{\rm kin}$ slices from a 3 dimensional grid of $\Bar{\sigma}^2$ for SO (\textit{top}), SiO (\textit{middle}) and CS (\textit{bottom}). Contours of constant $\Bar{\sigma}^2$ = 9 (darker); 25 (lighter) on a column density vs. \ce{H2} density grid for detected lines are shown. The background shades show the optical depths $\tau = 0.05$ (darker region) and 0.1 (lighter region) for all species. These optical depths correspond to the maximum optical depth out of all the lines modeled.}
        \label{fig:others_int_model}
    \end{figure*}

    {Calibration uncertainties were accounted for with a 20\% error for each line intensity. Figure \ref{fig:others_int_model} shows the projection of constant $\Bar{\sigma^2}$ on the column density and spatial density plane for a given kinetic temperature.}

    \begin{figure*}[!h]
        \centering
        \parbox{0.25\textwidth}{
            \includegraphics[width=0.25\textwidth]{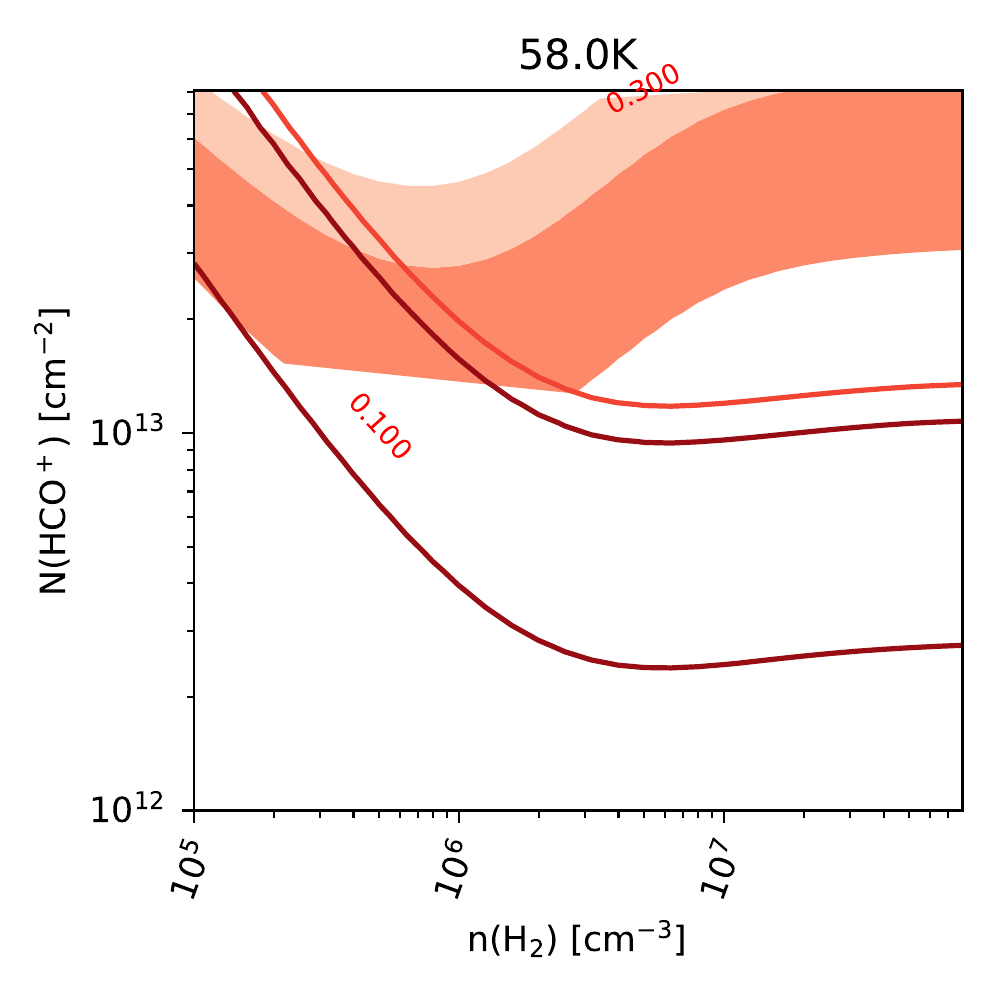}}
        \begin{minipage}{0.25\textwidth}
            \includegraphics[width=\textwidth]{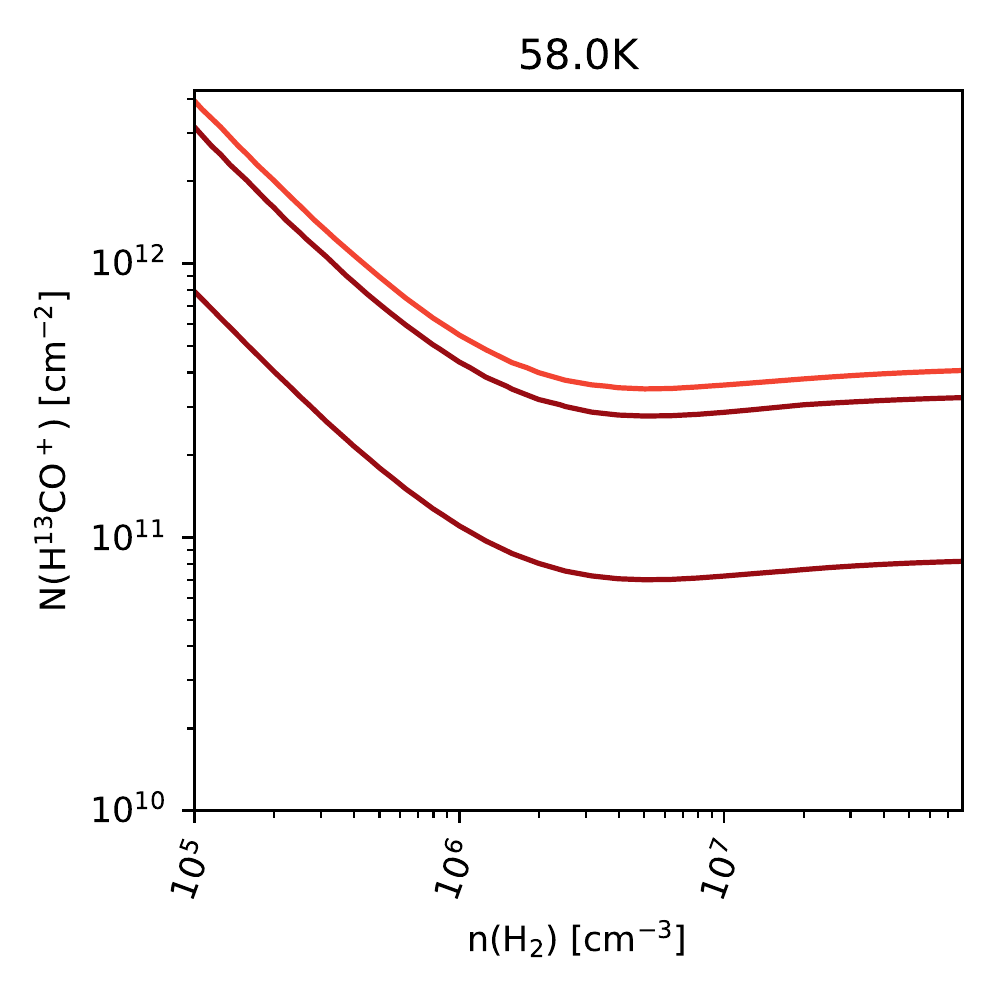}
        \end{minipage}
        \parbox{0.25\textwidth}{
            \includegraphics[width=0.25\textwidth]{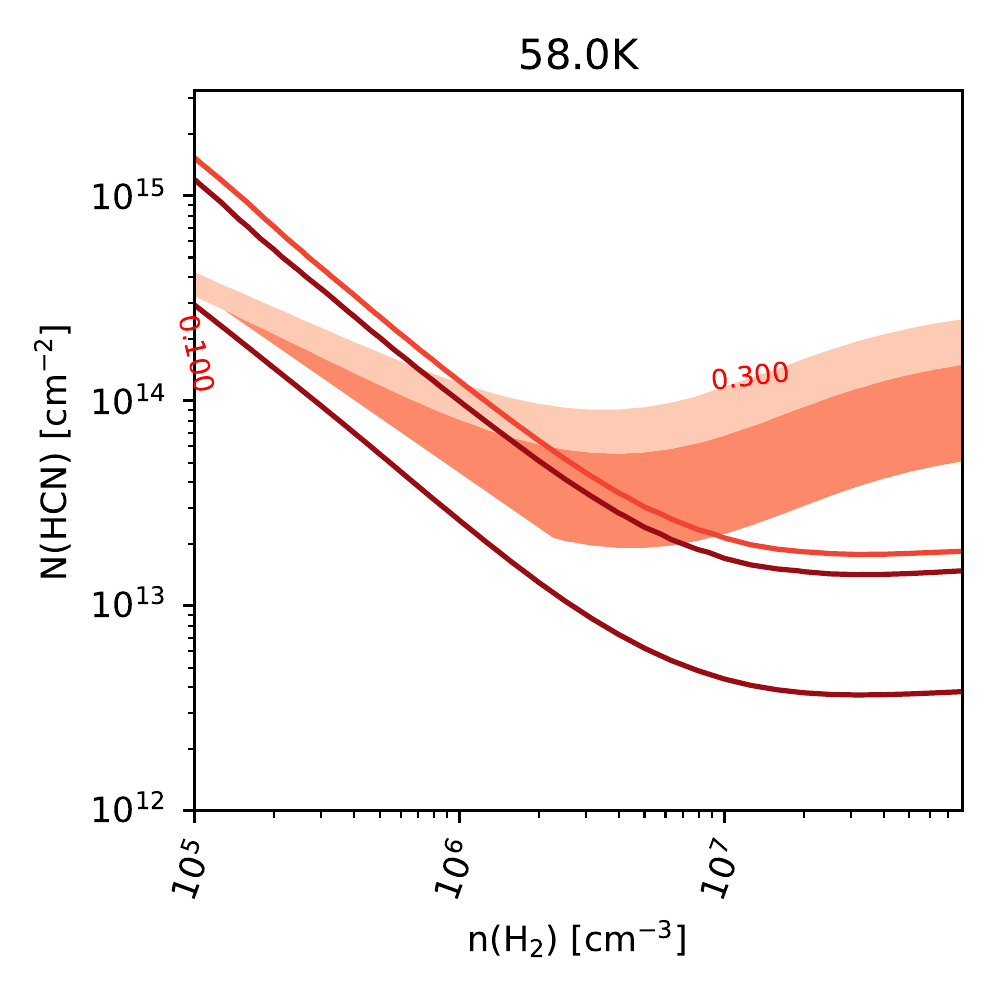}}
        \begin{minipage}{0.25\textwidth}
            \includegraphics[width=\textwidth]{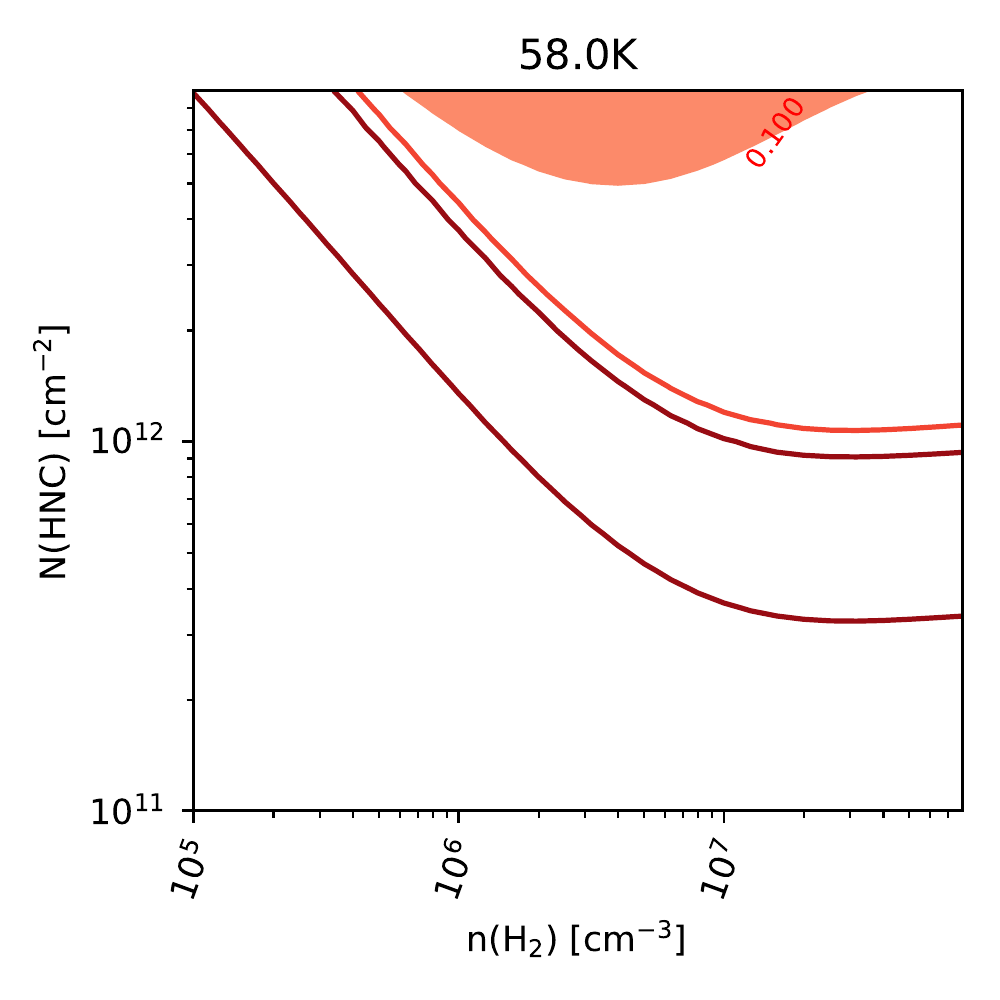}
        \end{minipage}
        \caption{Results of modeling individual line intensities with RADEX on a 2D grid of column densities and spatial densities for a constant $T_{\rm kin}$. Contours of constant $\Bar{\sigma}^2$ = 9 (darker); 25 (lighter) for detected lines are shown. The background shades show the optical depths $\tau = 0.05$ (darker region) and 0.1 (lighter region) for the corresponding species.}
        \label{fig:others_int_model_2}
    \end{figure*}

\end{appendix}

\end{document}